\documentclass[pra, twocolumn, showpacs, floatfix, superscriptaddress]{revtex4}
\usepackage{enumerate}
\usepackage{bbm}
\usepackage{bm}
\DeclareBoldMathCommand{\bfxi}{\xi}

\newcommand{\NewDelta}{\Delta}
\newcommand{\Newdelta}{\delta}
\newcommand{\modeindex}{\alpha}
\newcommand{\TMabbrev}{$p$}
\newcommand{\TEabbrev}{$s$}

\renewcommand{\Im}{{\rm Im}}
\usepackage{graphicx, xcolor}
\usepackage{epsfig}
\usepackage{pdfsync, hyperref}
\usepackage{amsmath}
\usepackage{wrapfig}

\begin{document}

    \title{Mode structure and polaritonic contributions to the\\ 
    Casimir effect in a magneto-dielectric cavity}
    \author{Harald R. Haakh}
\email{harald.haakh@mpl.mpg.de}
 \affiliation{
 Institut f\"ur Physik und Astronomie, Universit\"at Potsdam, 
 14476 Potsdam, Germany.
 }
\affiliation{
Max Planck Institute for the Science of Light, 
91058 Erlangen, Germany.}

\author{Francesco Intravaia}
 \affiliation{
 Institut f\"ur Physik und Astronomie, Universit\"at Potsdam, 
 14476 Potsdam, Germany.
 }
 \affiliation{Theoretical Division, Los Alamos National Laboratory, Los Alamos, NM 87545, USA.}
\affiliation{School of Physics and Astronomy, University of Nottingham, Nottingham NG7 2RD, United Kingdom.}

\date{19 August 2013}
    

\begin{abstract}
\noindent  We present a full analysis of the mode spectrum in a cavity formed by two parallel plates, one of which is a  magneto-dielectric, e.g. a metamaterial, while the other one is metallic, and obtain dispersion relations in closed form.
The optical properties of the cavity walls are described in terms of realistic models for the effective permittivity and the permeability.
Surface polaritons, i.e. electromagnetic modes that have at least partly an evanescent character, are shown to dominate the Casimir interaction at small separations. We analyze in detail the \TEabbrev -polarized polaritons, which are a characteristic feature of a magneto-dielectric configuration, and discuss their role in the repulsive Casimir force.
\end{abstract}

\maketitle

\section{Introduction}
\label{sec:introduction}
From the perspective of macroscopic electrodynamics, the  electromagnetic (e.\,m.) response of a material is described by its dielectric function $\epsilon(\omega)$ and magnetic permittivity $\mu(\omega)$. As both are functions of frequency, there is a panoply of dispersive phenomena \cite{Jackson75,Landau80}.
Many systems in nature and technology do not show any magnetic properties (especially at optical frequencies) and $\mu(\omega)$ can be set equal to unity. 
Recently, however, micro- or nanostructured composite media consisting of metals and/or dielectrics known as `metamaterials' or `plasmonic materials' have  attracted interest due to their unusual electric and, in particular, their magnetic properties, as well as their implications and applications in nanotechnology \cite{Zheludev10}. Metamaterials involving e.g. intrinsic magnetic composites \cite{Yannopapas09} or superconductors \cite{Ricci05,Du08,Chen10} allow for applying quantum phenomena like the Josephson effect \cite{Du08} for tayloring the material response to e.\,m.\ fields. 
In fact, metamaterials have become known for giving rise to phenomena such as negative refraction and superlensing \cite{Pendry00,Zheludev10} and have attracted much interest lately in the context of cloaking \citep{Leonhardt06, Pendry06, Cai07, Ergin10, Chen10a}.   
The peculiar optical features of nanostructured materials are greatly due to polaritonic modes \cite{Maier07}, i.e. collective excitations of matter coupled to the e.\,m.\ field.
While bulk polaritons are solutions of the wave equation in the bulk, surface polaritons, that are (at least partially) confined to an interface, 
dominate the e.\,m.\ near field.
The high degree of control  that can be achieved over the surface-mode (plasmon-)polariton spectrum was namegiving for the field of plasmonics and is fundamental for the aforementioned applications of metamaterials \cite{Pendry04}. 

It is therefore not surprising that researchers have also tried to use the optical properties of these new materials to modify dispersion interactions. An important example is the Casimir effect, that leads to forces between electrically neutral bodies at microscopic distances due to vacuum fluctuations of the e.\,m.\ field and matter polarization and magnetization \cite{Intravaia12b}.  Theoretical investigations have shown that a magnetic response can substantially modify the strength of the Casimir force \cite{Casimir48,Boyer74}. Magneto-di\-elec\-tric metamaterial configurations have therefore increasingly attracted attention and some work has been carried out to assess their potentials and limitations \cite{Henkel05, Lambrecht08, Pirozhenko08, Rosa08a,Rosa08,Rahi10a,Silveirinha10}. 

Here we study the case of a mixed cavity consisting of one metallic and one magneto-dielectric (meta)material plate, using a model that captures important features of a realistic system, and focus on the eigenmodes of the macroscopic e.\,m.\ field.
This approach has the merit of unveiling the underlying physical phenomena and points out delicate balances. 
We will concentrate on the role of  the surface polaritons, which are known to play a fundamental role in Casimir physics \cite{VanKampen68,Gerlach71,Intravaia05, Intravaia07, Haakh10}. Related work has shown that surface-mode control via nanostructured surfaces strongly affects the Casimir force \cite{Intravaia12, Guerout13}.
Cavities involving magneto-dielectrics give rise to a whole `zoo' of polaritonic modes, which has no equivalent in configurations involving non-magnetic media only. 
By systematically studying the mode structure, we extend previous treatments \citep{Lambrecht08, Pirozhenko08} and include also mode continua present in the mixed cavity.
We show that the polaritonic modes play a fundamental role in a repulsive Casimir interaction and address how other modes that belong to different frequency domains influence this phenomenon.
Our work also paves the way towards a modal analysis of the role of magneto-dielectric (meta)materials in other related fields like near-field heat transfer \cite{Volokitin07}, where polariton modes provide additional nonradiative channels, i.e.  enhanced heat transfer \cite{Joulain10, Francoeur11, Guerout12}. 

In this paper,
we will first discuss the mode structure in a mixed metal/magneto-dielectric cavity and provide the exact solutions for the surface-mode dispersion relations.
We then assess in Sec.\,\ref{sec:Casimir} their relevance for the Casimir effect, providing asymptotic expressions at short distances, where the Casimir force has a mainly polaritonic character.
Details of the calculations are given in the Appendices.
\section{Mode anatomy of a mixed cavity}
\label{sec:mode_anatomy}

\subsection{Description of the cavity}
\label{sec:bulk}

\begin{figure}[t!]
\centering
\includegraphics[width=\columnwidth]{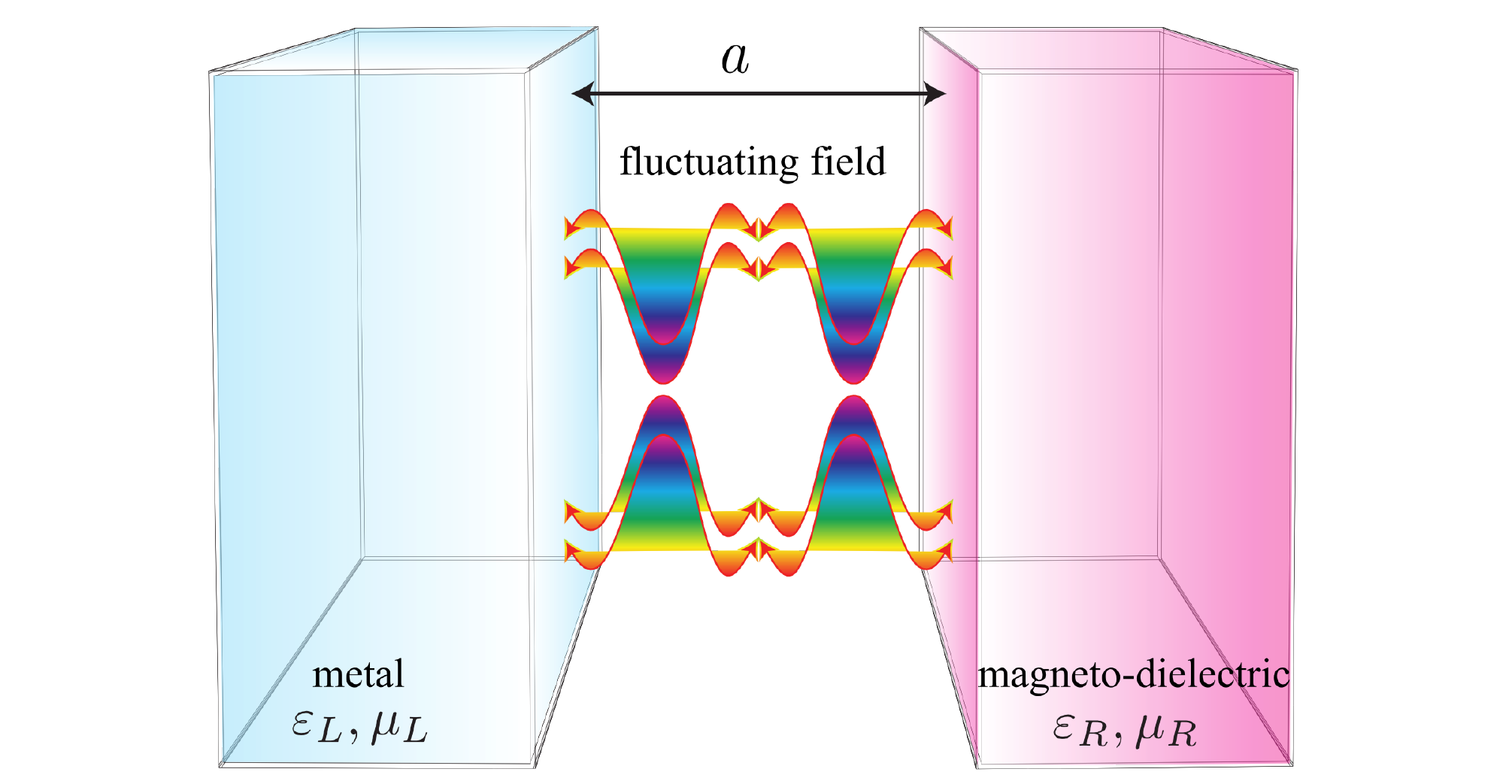}
\caption{(Color online) Mixed cavity formed by a metal surface and a magneto-dielectric (meta)material one, separated by a distance $a$.}
\label{fig:cavity}
\end{figure}
We start our analysis with a discussion of the mode structure of the e.\,m.\ field in a cavity formed by two plane parallel slabs at a distance $a$, one of which is made by a magneto-dielectric (right plate, index $R$) while the other is a normal metal (left plate, index $L$), as depicted in Fig.\,\ref{fig:cavity}.
 
The optical properties of the cavity are modeled in terms of an effective permittivity and permeability of each of the two media. For the magneto-dielectric plate we are going to use the local Lorentz-Drude model 
\begin{align}
\epsilon_{R}=1+\frac{{\chi_\epsilon} \omega_{r}^{2}}{\omega_{r}^{2}-\omega^{2}},\quad
 \mu_{R}&=1+\frac{{\chi_\mu} \omega_{r}^2}{\omega_{r}^{2}-\omega^{2}},
\label{eq:model_metamaterial}
\end{align}
which may describe both intrinsic magneto-dielectrics or -- within the effective medium approach -- metamaterials \cite{Rosa08a}.
We have assumed that both $\epsilon_R$ and $\mu_R$ have a common resonance frequency $\omega_r$. This simplified model catches most of the important features of the system, maintaining an accessible mathematical complexity.
The dimensionless coefficients ${\chi_\epsilon}>0, \chi_\mu$ give the static susceptibilities.
The sign of ${\chi_\mu}$ distinguishes between diamagnetic (negative) and paramagnetic (positive) media.
As effective media, metamaterials that contain normally conducting nanostructures can be either diamagnetic \cite{Grigorenko05} or paramagnetic \cite{Pendry99, Rosa08}.
However, the magnetic coefficient ${\chi_\mu}$ must satisfy some physical constraints.
Most media fulfill the Bohr-Van Leeuwen theorem \cite{Leeuwen21} so that $|\mu(\omega \to 0)| \approx 1$ (`classical' materials).
As this implies $|{\chi_\mu}|\ll1$, any relevant magnetic activity  in these systems occurs at frequencies close to the resonance where $|\mu(\omega)| \gg 1$.
For these parameters, the model of Eq.\,\eqref{eq:model_metamaterial} can be expressed in the form 
\begin{align}
\mu_R = 1 + \chi_\mu + \frac{\chi_\mu \omega^2}{\omega_r^2  -\omega^2}
\approx 1+ \frac{\chi_\mu \omega^2}{\omega_r^2 - \omega^2}
\end{align}
which is equivalent to the model used in Refs.\,\citep{Pendry99, Rosa08a}.

\begin{figure}[t!]
\begin{center}
 \includegraphics[width=\columnwidth]{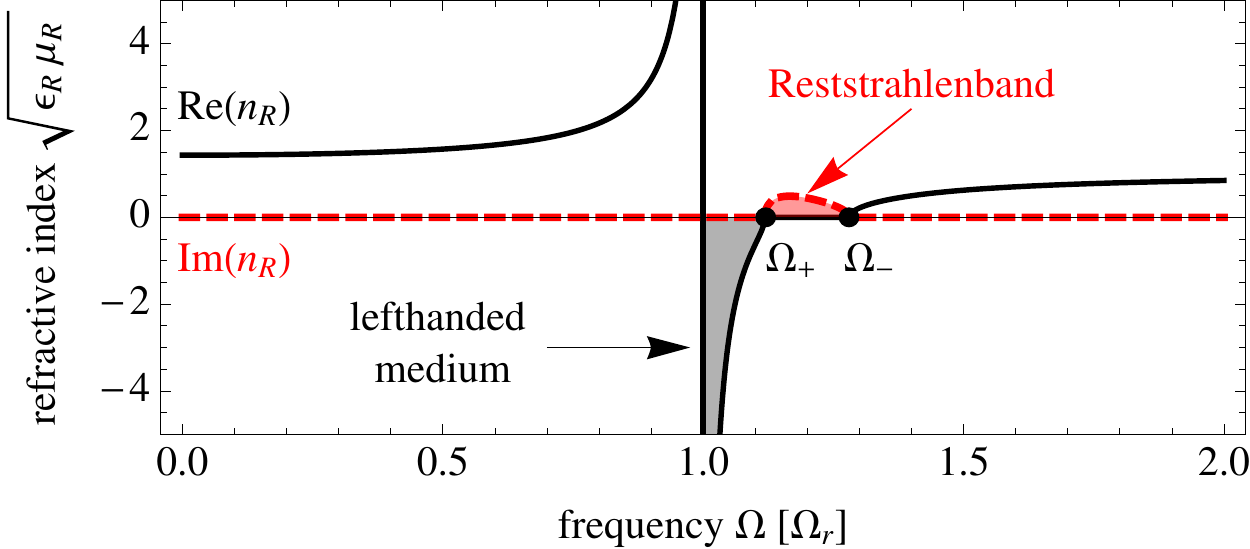}
   \caption{
   (Color online) 
Real and imaginary part of the effective refractive index $n_R = \sqrt{\epsilon_R \mu_R}$ of the magneto-dielectric medium as a function of frequency.  The material parameters are ${\chi_\epsilon} = 0.25$, ${\chi_\mu} = 0.64$.
The lefthanded regime  [iii) of Fig.\,\ref{fig:bandlimits}], where $n_R < 0$and the Reststrahlenband [regime iv)] are indicated by the shaded areas, limited by the frequencies $\Omega_r$ and $\Omega_\pm(K=0)$. Note that the the refractive index is intimately connected to the propagation constant at $K=0$ [cut along the left border of Fig.\,\ref{fig:bandlimits}a)], where $\imath \kappa_R(K=0) = n_R \Omega$.
}
    \label{fig:index}
\end{center}
\end{figure}

When, in contrast, `nonclassical' media such as superconductors \cite{Du06} or strong ferromagnets (paramagnetic materials) \cite{Yannopapas09} are used as constituents of a metamaterial, a strong static magnetic response is actually possible.
Our model therefore covers a wide range of values of ${\chi_\mu}$. In the numerical calculations below we focus on this last case  of a nonclassical material and use ${\chi_\epsilon} = 0.25$, ${\chi_\mu} = 0.64$ (paramagnetic magneto-dielectric) to describe the features of our system. For these values, the magneto-dielectric behaves like a lefthanded material at certain frequencies, i.e. it features a negative index of refraction  (Fig.\,\ref{fig:index}). For the diamagnetic case ($\chi_{\mu}<0$), calculations follow along the lines of the present work.
Nonwithstanding the exact choice of a composite medium, one should keep in mind that the previous effective medium description is only valid for e.\,m.\ fields that cannot resolve the internal structure of the metamaterial, i.e. to wavelengths in the optical range or above. For intrinsic magneto-dielectrics there is, of course, no such constraint.

The left plate of our cavity is metallic and may be reasonably approximated by the lossless (`plasma') response 
\begin{equation}
\epsilon_{L}=1-\frac{\omega_{r}^{2}}{\omega^{2}},\qquad  \mu_L = 1~.
\label{eq:metal}
\end{equation}
%
To simplify the forthcoming calculations, we  assume that the plasma frequency of the metallic plate coincides with the resonance frequency $\omega_r$ of the Lorentz-Drude model. Some implications of this assumption are discussed in App.\,\ref{app:tangent}.
To give an idea of the physical scales of the system we recall that a silver mirror is characterized by a plasma frequency $\omega_r = \Omega_r /a= 1.37\times 10^{16}\,\rm{s^{-1}}$ corresponding to a plasma wavelength $\lambda_r = 2\pi c / \omega_r = 137\,\rm nm$.

The cavity length $a$ provides an inherent scale of the system by which frequencies and wave vectors can be measured, so that it is convenient to use dimensionless variables throughout the calculation ($\hbar = c = 1$).
Having two parallel plates, the symmetry of the system indicates the frequency $\omega$ and the components of the wave vector parallel to the interfaces $\mathbf{k}_\parallel$
as relevant quantities. We thus define the dimensionless quantities $\Omega= \omega a$, $K= |\mathbf{k}_\parallel| a$ and similarly for any other relevant frequency or momentum. 

 In the following, we perform a full mode analysis of a cavity of media described by Eqs.\,\eqref{eq:model_metamaterial} and \eqref{eq:metal}.
We are interested in functions of the form $\Omega(K)$ to describe mode dispersion relation and band limits. As it turns out, many of these functions are solutions of transcendental equations and do not give rise to simple expressions. We deal with this issue by introducing a parametrization $\Omega(K)\equiv \{\Omega(z),K(z)\}$ with $z=K^{2}-\Omega^{2}$ so that we can interchangeably use $\Omega(z)$ or $\Omega(K)$. Eventually, we find these curves as solutions $\epsilon_L= \epsilon_{\modeindex}(z)$  of polynomial equations. From there and Eq.\,\eqref{eq:metal}, we find $\Omega(z)$ according to
\begin{equation}
\label{eq:omega_from_epsilon}
\Omega_{\modeindex}(z)=\Omega_{r} \frac{1}{\sqrt{1-\epsilon_{\modeindex}(z)}}~.
\end{equation}
Detailed calculations are given in the Appendices and the results are collected in Table\,\ref{tab:dispersion}.
Note that the sign of $z$ also distinguishes the modes according to their physical characteristics: As $z=0$ defines the light cone, negative values of $z$ correspond to propagating waves inside the cavity, while positive values describe evanescent waves.

\begin{figure*}[p!]
\begin{center}
\raisebox{0cm}{a)} \includegraphics[width=.45\textwidth]{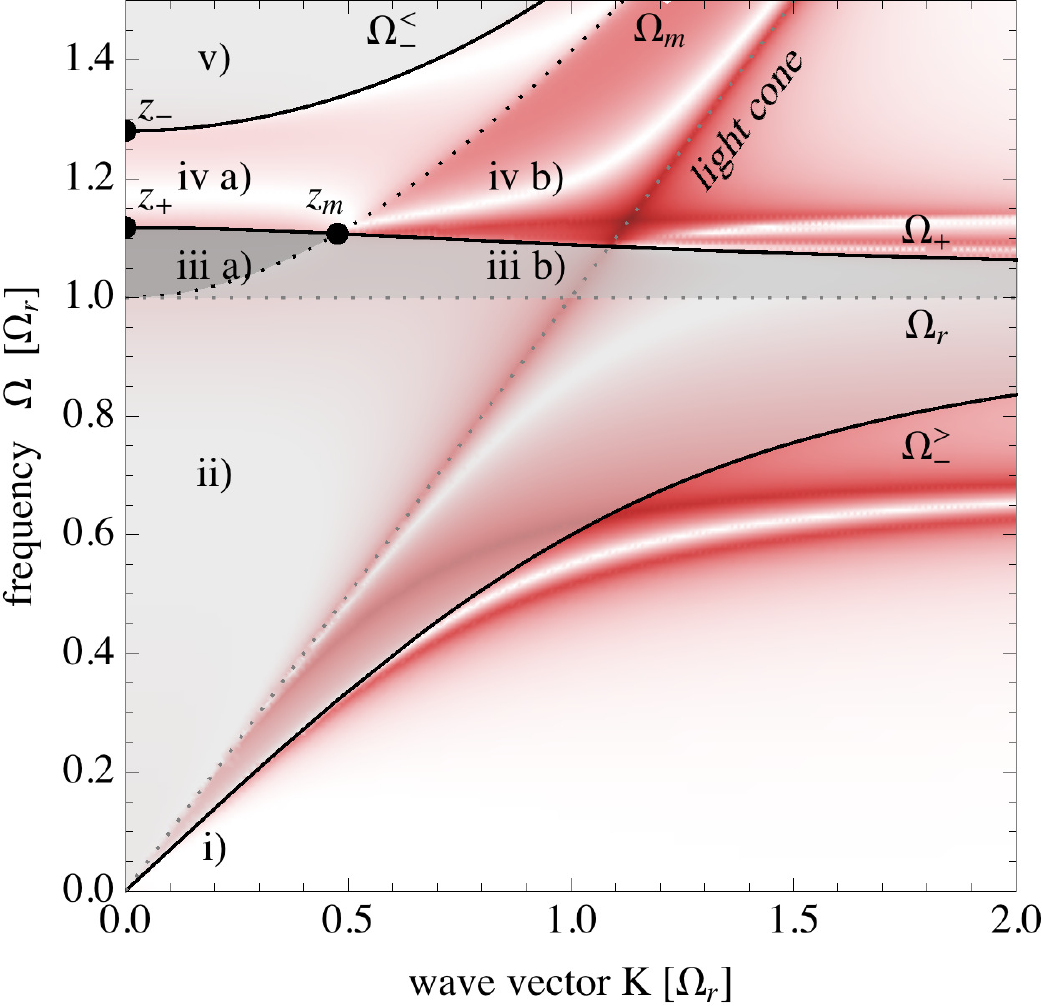}
\hspace{5ex}
\raisebox{0cm}{b)}\includegraphics[width=.45\textwidth]{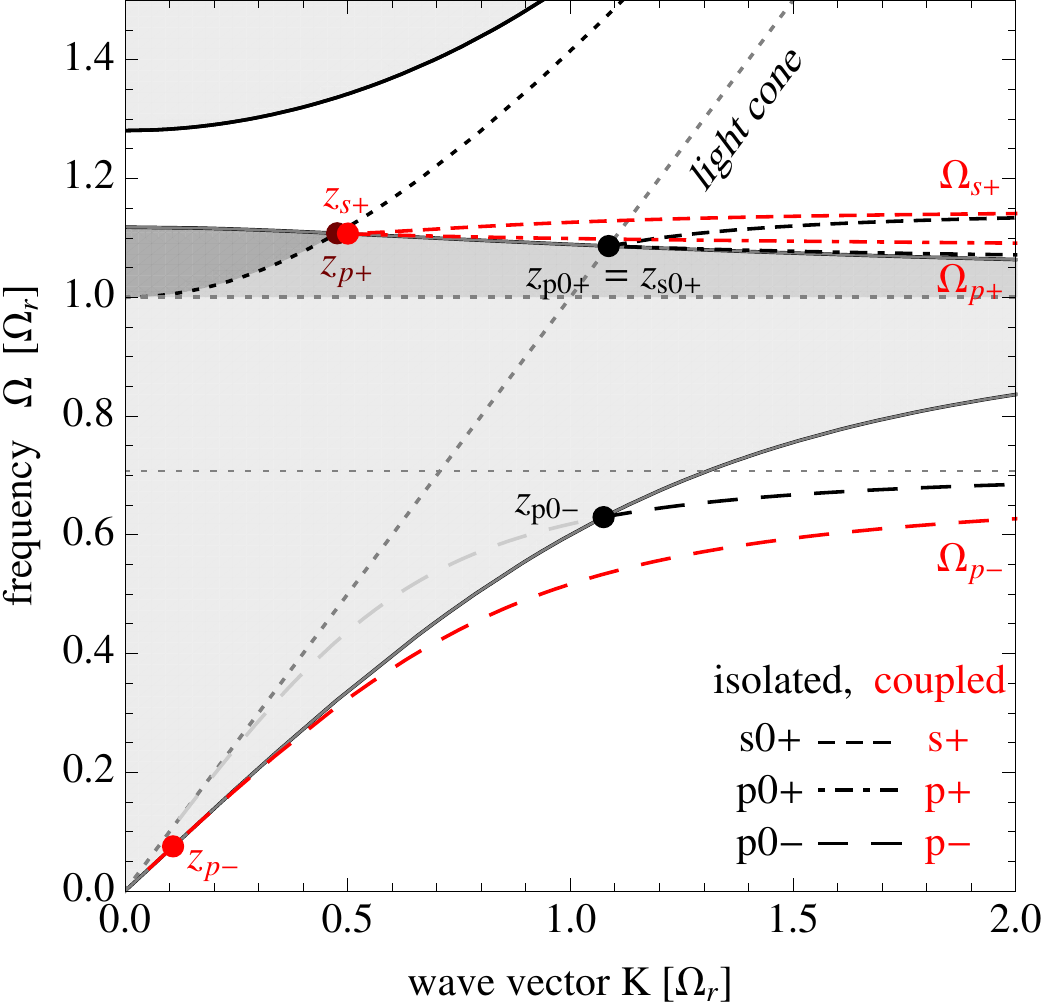}
\vspace{-0.3cm}
   \caption{
   (Color online) 
a) Band structure inside the cavity walls. Inside the magneto-dielectric plate, evanescent fields occur in zone i) and in the Reststrahlenband iv).
Light gray shading indicates zones ii) and v) where the plate behaves as a righthanded medium while darker shading denotes the lefthanded response in zone iii).
In the metal plate, the dotted curve $\Omega_{m}(K)$ provides a lower frequency boundary to propagating bulk modes.
This includes a portion iii\,a) of the lefthanded band and a portion iv\,a) of the Reststrahlenband.
The color-coded background gives a numerical evaluation of the Lifshitz energy density (absolute value), 
cf. Eq.\,\eqref{eq:Lifshitz}.
Darker color indicates a higher mode density and mirrors the mode structure of panel b).
The numerical calculations were performed for a cavity length $2 \pi a / \lambda_r = 0.1$ and material parameters ${\chi_\epsilon} = 0.25$, ${\chi_\mu} = 0.64$.
\newline
b)  Dispersion relations of isolated surface modes (black curves) and of the  coupled cavity polaritons (red curves),  see also Table \ref{tab:dispersion}. 
Inside the magneto-dielectric bulk continuum [zones ii) and iii)], the isolated polaritonic modes connect to broadened resonances.
}
    \label{fig:bandlimits}
\end{center}
\end{figure*}

{
\begin{table*}[p!]
\small
\begin{tabular}{llll}
\hline \hline
function &symbol&  (parametric) representation & end point\\[1ex]
\hline\\[-2ex]
\parbox[c]{1.7cm}{\raggedright metal band limit} 
&$m$ 
&$
z = -\Omega_r^2\quad \Leftrightarrow\quad \Omega_m(K) = \sqrt{K^2 + \Omega_r^2} 
$
&
$z_m = -\Omega_r^2$
\\[2ex]

\parbox[c]{1.7cm}{\raggedright mag.-diel.\\band limits} 
& $\pm$
&
$\epsilon_{\pm}(z)=
\displaystyle \frac{\Newdelta }{\frac{\NewDelta }{2} \pm \sqrt{\left(\frac{\NewDelta }{2}\right)^{2}+z \Newdelta}}
$
&
\parbox[c]{1.7cm}{
$z_-=- \Omega_r^2 \tilde \epsilon$\\
$z_+=- \Omega_r^2 \tilde \mu$}

\\[2ex]
%
\hline\\[-2.5ex]
\parbox[c]{1.7cm}{\raggedright isolated mag.-diel. polariton} 
&$s0+$ & 
$\displaystyle\epsilon_{s0+}(z)=\frac{\left(\frac{\Newdelta }{z{\chi_\mu}}+{\chi_\mu}\right)}{\left(\tilde{\mu}+\frac{\NewDelta }{2 z {\chi_\mu}}\right)
+\sqrt{\left(\tilde{\mu}+\frac{\NewDelta }{2 z {\chi_\mu}}\right)^{2}+\frac{1-\tilde{\mu}^{2}}{{\chi_\mu}}\left(\frac{\Newdelta }{z{\chi_\mu}}+{\chi_\mu}\right)}}
$
&
$z_{s0+}=0$
\\
\parbox[c]{1.7cm}{\raggedright isolated mag.-diel. polariton} 
& $p0+$ &
$\displaystyle 
\epsilon_{p0+}(z)=\frac{\left(\frac{\Newdelta }{z {\chi_\epsilon}}+{\chi_\epsilon}\right)}{\left(\tilde{\epsilon}+\frac{\NewDelta }{2z{\chi_\epsilon}}\right)
+\sqrt{\left(\tilde{\epsilon}+\frac{\NewDelta }{2z{\chi_\epsilon}}\right)^{2}+\frac{1-\tilde{\epsilon}^{2}}{{\chi_\epsilon}}\left(\frac{\Newdelta }{z{\chi_\epsilon}}+{\chi_\epsilon}\right)}}
$
&
$z_{p0+}=0$
\\

\parbox[c]{1.7cm}{\raggedright isol. metal plasmon} 
&$p0-$ &
$\displaystyle \epsilon_{p0-}(z)= - \sqrt{1 + \Omega_r^2/z}
\quad \Leftrightarrow \quad \textstyle \Omega_{p0-}(K) = \sqrt{K^2+ (\frac{\Omega_r}{\sqrt{2}})^2-\sqrt{K^4+(\frac{\Omega_r}{\sqrt{2}})^4}}$
&
$z_{p0-}: \epsilon_{-}(z) = \epsilon_{p0-} (z)$
\\[2ex]
\hline
\\[-2.5ex]
\parbox[c]{1.7cm}{\raggedright cavity polariton} 
&
 $s+$
&
$\displaystyle
\epsilon_{s+}(z)=\frac{\left(\frac{\Newdelta }{z {\chi_\mu}}+{\chi_\mu}C_s^{2}\right)}
{\left(\tilde{\mu}C_s^{2}+\frac{\NewDelta }{2z {\chi_\mu}}\right)
+\sqrt{\left(\tilde{\mu} C_s^{2}+\frac{\NewDelta }{2z{\chi_\mu}}\right)^{2}+\frac{1-\tilde{\mu}^{2}C_s^{2}}{{\chi_\mu}}\left(\frac{\Newdelta }{z {\chi_\mu}}+{\chi_\mu}C_s^{2}\right)}}
$
&

$\begin{array}{l}
z_{s+}: 
\text{tangent point}
\\
\text{Eq.\,\eqref{eq:def_zsplus}}
\end{array}
$
\\
\raisebox{1.5ex}{
\parbox[c]{1.7cm}{\raggedright cavity \\$+$ polariton \\ $-$ plasmon} } 
&
$p\pm$
&
$\epsilon_{p\pm}(z)$: 

$\displaystyle
\displaystyle [(1+{\chi_\epsilon})\epsilon_{p\pm}-{\chi_\epsilon}]^{2}(1+\epsilon_{p\pm} T_{0}T_{s})^{2}=(\epsilon_{p\pm} T_{s}+T_{0})^{2}
\left(\epsilon_{p\pm}^2+\frac{\NewDelta  \epsilon_{p\pm}}{z}-\frac{\Newdelta }{z}\right)
$
&
$\begin{array}{l}
z_{p\pm}: 
\text{tangent point}
\\
\text{Eqs.\,\eqref{eq:def_zpplus}, \eqref{eq:def_zpminus}}
\end{array}
$
\\[-2ex]
&&\text{select solutions such that}: 
$\displaystyle \epsilon_{p+} \to \frac{\tilde{\epsilon}-1}{\tilde{\epsilon}+1}~, \quad
\epsilon_{p-}\to -1$ as  $z \to \infty$.
\\[2ex]

\hline \hline
\end{tabular}
\vspace{-0.3cm}
\caption{Band limits and isolated and cavity polariton dispersion relations. Curves $\Omega(K)$ are obtained from Eq.\,\eqref{eq:omega_from_epsilon} and the relation $z= K^2 - \Omega^2$.
We use the abbreviations 
\mbox{$\Newdelta =\Omega^{2}_{r}{\chi_\mu}\,{\chi_\epsilon}$}, 
 \mbox{$\NewDelta =\Omega^{2}_{r}({\chi_\mu}\,{\chi_\epsilon}+{\chi_\mu}+{\chi_\epsilon})$} 
 (both are positive at the values $\chi_{\epsilon}$ and $\chi_{\mu}$ chosen in the numerics)
  as well as $\tilde{\mu} = 1 +{\chi_\mu}$ and $\tilde{\epsilon} = 1 + {\chi_\epsilon}$. 
The quantities $C_s, T_s, T_0$ are defined in Eqs.\,\eqref{eq:coupled}, \eqref{eq:coupled2}.
}
\label{tab:dispersion}
\end{table*}
}

\subsection{Wave propagation in the bulk}
\label{sec:bands}

For a full understanding of the mode structure in the cavity it is necessary to review how e.\,m.\ fields behave inside the slabs.
This is characterized by the purely real or purely imaginary propagation constant inside the left ($L$) or right ($R$) mirror
\begin{align}
\label{eq:def_kappa}
\kappa_{i}^{2}=z-\Omega^{2}(\epsilon_{i}\mu_{i}-1)
,\quad i = L,R~.
\end{align}
For example, in the case of a wave inciding from vacuum onto the right (magneto-dielectic) mirror, $\mathrm{Re}[\kappa_{i}]\ge0$ will correspond to a wave that decays in the same direction inside the material (evanescent wave), while $\mathrm{Im}[\kappa_{i}]\le0$ indicates a propagation through the medium ($\propto e^{-\kappa_i}$). 
In vacuum, the propagation constant is $\kappa_0^{2} = z$ with $\Im[\kappa_0]<0$ for a propagation from left to right.

In each medium, propagating and evanescent behavior is separated by $\kappa_i = 0$. Inside the metallic slab the closed solution of $\kappa_L^2=0$ is easily found and it is given by  $\Omega_{m}(K)=\sqrt{\Omega_{r}^{2}+K^{2}}$ 
 \cite{Intravaia07,Haakh10}.
It separates two different regimes in the $\Omega$-$K$-plane, as indicated in Fig.\,\ref{fig:bandlimits} \cite{Intravaia07,Haakh10}:
\begin{enumerate}[a)]
\item 
\emph{metal bulk (plasmon) modes} that can propagate inside the metal ($\kappa_L^{2} <0$) and cover the region of frequencies above $\Omega_{m}(K)$ , and
\item
\emph{evanescent fields} inside the metal plate, characterized by $\kappa_L ^{2} >0$ at frequencies below $\Omega_{m}(K)$.
\end{enumerate}

As it appears from  Fig.\,\ref{fig:bandlimits}, things are more involved inside the magneto-dielectric slab, that may behave as a left-handed material, depending on the sign of the  propagation constant \cite{Veselago06}. 
The two solutions of  $\kappa_R ^{2}= 0$ define the \emph{magneto-dielectric band limits} $\Omega_{\pm}(K)$ (Fig.\,\ref{fig:bandlimits}).
The corresponding parametric expressions 
are collected in Table \ref{tab:dispersion} and details of the calculation are given in App.\,\ref{app:band_limit}.

Both  solutions tend to the same frequency $\Omega_{\pm} \to \Omega_r$ as $K \to \infty$.
The curve $\Omega_+$ crosses the light cone
and intercepts the $K=0$ axis at a value $\Omega_+(z_+)$.
The solution $\Omega_{-}$ splits up into two branches, 
depending on the sign of $z$:
The frequency $\Omega_{-}^{>}(z)$ corresponding to $z>0$  goes to zero with the wave vector $K$ and becomes tangential to the light cone.
The second branch $\Omega_{-}^{<}$ (values $z<0$) approaches a the lower magneto-dielectric band limit that behaves as a medium light cone of index $n_R = \sqrt{(1 +\chi_\mu) (1 + \chi_\epsilon)}$ at low frequencies (cf. Figs.\,\ref{fig:index} and \ref{fig:bandlimits}).


Together with the resonance frequency $\Omega_r$ these curves define the zones shown in Fig.\,\ref{fig:bandlimits}a) and characterize the propagation of the electromagnetic field inside the magneto-dielectric:
%
\begin{enumerate}[i)]
\item
$\Omega<\Omega_{-}^{>}(K)$  the e.\,m.\ field is evanescent inside the medium
($\kappa_{R}^{2} >0$). 
\item
$\Omega_{-}^{>}(K)<\Omega<\Omega_{r}$ and $\Omega_{-}^{<}(K)<\Omega$ (second branch) the magneto-dielectric is \emph{righthanded}: for an incident wave from the left the e.\,m.\ field propagates inside the medium in the same direction ($\Im[\kappa_{R}]  <0$).
\item
$\Omega_{r}<\Omega<\Omega_+(K)$ the magneto-dielectric behaves like a \emph{lefthanded} material \cite{Veselago06, Pendry00}: the e.\,m.\ field is still propagating inside the medium, but the propagation direction is opposite to the incident wave. In the case of an incident wave from the left this means $\Im[\kappa_{R}]>0$. 
\item
$\Omega_+{}(K)<\Omega<\Omega_{-}^{<}(K)$ defines the so-called \emph{Reststrahlenband}. Here, the magneto-dielectric completely reflects an incident propagating field e.\,m.\ field, ($|r_{R}^{s,p}|=1$) allowing only for transmitted evanescent waves with $\kappa_{R}^{2}>0$. 
\item
$\Omega_{-}^{<}(K)<\Omega$ the magneto-dielectric behaves again like a usual \emph{righthanded} material and supports propagating waves with ($\kappa_{R}^{2}<0$). This the equivalent of the \emph{bulk modes} region encountered in the metal plate for frequencies above the plasma frequency.
\end{enumerate}
Together, these regimes provide a general classification of the cavity modes, which we will now study in detail. Note that the previous discussion has unambiguously defined the signs of the square root arising from Eq.\,\eqref{eq:def_kappa} and of the refractive index, see Fig.\,\ref{fig:index}. These signs demand special attention in numerical evaluations.

\subsection{Polaritons on a single surface}
\label{sec:isolated_plasmons}

We now turn to the surface polariton dispersion relations (SPDRs) for a single interface.
In the case of semi-infinite plane mirrors, all necessary information can be obtained from the reflection coefficients and, eventually, from the surface impedances $Z_{i}^\sigma$ 
\begin{equation}
\label{eq:}
r^\sigma_{i}=\frac{1-Z^\sigma_{i}}{1+Z^\sigma_{i}}~, \quad i=L, R~,
\end{equation}
where $\sigma=s, p$ indicates the field polarization (TE or TM polarization, respectively).
Within the limits of the effective medium description, the response can be considered in the Fresnel (local)  limit, so that
$Z_{i}^{s}=\mu_{i}\sqrt{z}/\kappa_{i}$ and $Z_{i}^{p}=\epsilon_{i}\sqrt{z}/\kappa_{i}$ for the 
left or right surface ($i = L, R$), respectively.
From the poles of the reflection coefficients $r_i^\sigma$ of the individual plates we find the dispersion relations of surface-mode excitations. This corresponds to the solutions of
\begin{equation}
\label{eq:isolated}
\mu_{R}\sqrt{z}=-\kappa_{R}~ \text{($s$-pol.)},\quad
\begin{cases}
&\epsilon_{L}\sqrt{z}=-\kappa_{L}\\
&\epsilon_{R}\sqrt{z}=-\kappa_{R}
\end{cases}~ \text{($p$-pol.)}.
\end{equation}
There is only a single \TEabbrev -polarized surface polariton on the magneto-dielectric mirror, the dispersion relation of which is given by $\Omega_{s0+}(K)$. [Here, the mode index denotes  an \TEabbrev-polarized, isolated (index `0') mode  of frequency larger than $\Omega_r$ (index `$+$'). Other modes will be labeled similarly.]
In contrast, each of the two isolated surfaces carries a \TMabbrev -polarized mode: the surface polariton $\Omega_{p0+}$  on the isolated magneto-dielectric and the well-known surface plasmon $\Omega_{p0-}$ on the metal surface. The explicit \mbox{SPDRs} are collected in Table \ref{tab:dispersion} (see App.\,\ref{app:SPDR} for details). 

As we see from Fig.\,\ref{fig:bandlimits}b), the two surface modes $\Omega_{s0+}$ and $\Omega_{p0+}$ of the magneto-dielectric end at the band limit $\Omega_+$. This happens because solutions of the previous equations exists only when $\kappa_{R}$ is a real number ($\kappa_R^2 >0$), i.e.  outside the transparent zones ii) and iii) of Fig.\,\ref{fig:bandlimits}a).
A further necessary condition is that $\epsilon_{R}, \mu_{R} <0$ so that $\Omega_{r}<\Omega< \sqrt{\Omega_{r}^{2}+\Omega^{2}}$.
This leaves the part of region iv) of the Reststrahlenband below the light cone as unique region where the isolated polariton can live (see also App. \ref{app:tangent} for more details). 

Right on the vacuum light cone $z=0$, where we find that 
$\epsilon_{s0+}(z=0)=\epsilon_{p0+}(z=0)= \epsilon_{+}(z=0)$. 
%
This confirms that the isolated polariton stops precisely at the intersection between the band limit $\Omega_+$ and the vacuum light cone, corresponding to $z_{s0+} = z_{p0+} =0$. 

Finally, on the metal side, there is  the well-known surface plasmon (polariton) mode \citep{Intravaia07,Haakh10}. This corresponds to the solution $\epsilon_{p0-}$ obtained from Eq.\,\eqref{eq:isolated} and given in Table \ref{tab:dispersion} along with the others. The dispersion curve $\Omega_{p0-}(z)$ crosses the magneto-dielectric's band limit given by $\Omega_{-}^{>}(z)$, see Fig.\,\ref{fig:bandlimits}b). The corresponding crossing point  $z_{p0-}>0$ can be obtained by solving the equation $\epsilon_{-}(z)=\epsilon_{p0-}(z)$.
For $0<z<z_{p0-}$ the metallic plasmon mode overlaps with the magneto-dielectric bulk continuum [region ii) of Fig.\,\ref{fig:bandlimits}a)], where $\kappa_{R}$ is imaginary (propagating waves inside the magneto-dielectric).
We are going to see how this overlap puts constraints on the coupled polaritons and affects the corresponding contribution to the Casimir energy.

\subsection{Cavity polariton modes 
}
\label{sec:cavity_modes}
%
When the metallic and the magneto-dielectric interface are brought close to each other the surface polaritons of each interface couple through the e.\,m.\ field and hybridize. As a result, there are three distinct cavity polariton modes, given by the solutions of
\begin{equation}
\label{eq:Dispersion}
1-r^\sigma_{L}(\Omega)r^\sigma_{R}(\Omega)e^{-2\sqrt{z}}=0.
\end{equation}
As before, the index $\sigma=s, p$ denotes the two polarizations. 
Simple algebraic manipulation of Eq.\,\eqref{eq:Dispersion} together with the Fresnel coefficients gives
\begin{align}
\label{eq:coupled}
\begin{cases}
\mu_{R}\sqrt{z}C_s =-\kappa_{R} \text{ ($s$-pol.)}\\
%
\epsilon_{R}\sqrt{z} C_p = -\kappa_R \text{ ($p$-pol.)}
\end{cases}
\text{ with }C_{\sigma}
=
\frac{1+T_{0}T_{\sigma}}{T_{\sigma}+T_{0}}
\end{align}
%
and where we have defined
\begin{equation}
\label{eq:coupled2}
T_{0}=\tanh(\sqrt{z}), \quad T_{s}=
\sqrt{\frac{z}{z+\Omega_{r}^{2}}}, 
\quad T_{p} = \epsilon_L T_{s}.
\end{equation}
Obviously, the large distance limit $T_0 \to 1 \Rightarrow C_\sigma \to 1$ recovers two of the previous expressions \eqref{eq:isolated}.
The third one follows in the limit $T_0 \to 1$, when both the numerator and the denominator of $C_p$ vanish.

Let us consider first the \TEabbrev -polarized excitations.
We show in App.\,\ref{app:SPDR} that in this case the
only physical solution is  $\epsilon_{s+}$, given in Table \ref{tab:dispersion}. 
The dispersion relation $\Omega_{s+}$ is given in Fig.\,\ref{fig:bandlimits}b) (red dash-dotted curve). 
At large values of the wave vector, the curve  coincides with its isolated counterpart $\Omega_{s0+}$.  However, it does not stop at the light cone any more
but is pulled into the Reststrahlenband until it touches the band limit $\Omega_+$ in a tangent point.
The corresponding value of $z$ is found in the interval  $-{\pi^2}/{4 }\le z_{s+}\le 0$ 
with a lower limit provided by  $z=-\Omega_r^2 = z_m$ at small distances (see App.\,\ref{app:SPDR}).

From the mathematical point of view, discrete eigenvalues of the wave equation cannot overlap with continuous ones. Indeed, we have seen that $\kappa_{R}$ or $\kappa_{L}$ is imaginary beyond the band limits so that mathematically the corresponding equation in Eq.\,\eqref{eq:coupled} cannot be fulfilled (see App. \ref{app:tangent}).
This implies the stopping of $\Omega_{s0+}$ on the light cone in the case of a single surface, where the extended medium is the vacuum, whereas in a cavity the limit to $\Omega_{s+}$ is provided by the magneto-dielectric branchcut $\Omega_+$ or - eventually - by the metal one $\Omega_m$.

In the case of \TMabbrev -polarized fields the two expressions for $\epsilon_{p\pm}$ 
are the solution of the fourth-order polynomial equation reported in Table \ref{tab:dispersion}.
Closed analytic forms of the solutions can be obtained but they are not very instructive.
Some important limits  are discussed in App.\,\ref{app:SPDR}. From Fig.\,\ref{fig:bandlimits}b) one can see that the dispersion curve $\Omega_{p+}$ approaches the isolated polariton at large values of $z$ and crosses the light cone before it finally becomes tangent to the band limit at a point  $\max \{z_m, -\pi^2/4 \}< z_{p+}<0$.
The qualitative similarity between the \TMabbrev -polarized mode $\Omega_{p+}$ and the \TEabbrev -polarized mode $\Omega_{s+}$ is obvious, yet the endpoints $z_{p+}$ and $z_{s+}$  do not coincide generally. The second solution, $\Omega_{p-}$, is the cavity counterpart of the metallic plasmon and it is a pure surface mode (entirely evanescent). As in the isolated case it takes a value $\Omega_r / \sqrt{2}$ as $z \to \infty$ but approaches the lower branch of the band limit $\Omega_{-}^{>}$ (rather than the vacuum light cone) at smaller values of $z$, stopping at a tangent point $z_{p-}\ge 0$, see Eq.\,\eqref{eq:def_zpminus}.

This underlines that the band limits $\Omega_{\pm}$ take the role of `medium light cones'. They are also branching points of the square roots involved in the reflection coefficients  and physically give rise to a continuum of bulk modes that covers the regions ii), iii), and v) of the $\Omega$-$K$-plane.
This can be seen very clearly from the numerical evaluation of the density of states in Fig.\,\ref{fig:bandlimits}a), where darker coloring indicates higher mode densities.
From the metal mirror, too, the mode continuum of bulk plasmon polaritons provides a similar contribution in the region a).
Note, that cavity walls of finite width  \cite{Schram73,Economou69} (or a finite quantization volume) would result in discrete modes inside the same frequency bands but would not alter greatly the physics. 


\section{Casimir interaction in a mixed cavity}
\label{sec:Casimir}
Now that we have achieved a good understanding of the mode structure of the cavity, we can evaluate the
polaritonic contribution to the Casimir energy for a plate area $A$ at zero temperature.
Knowing the mode-spectrum of the cavity, it is useful to invoke the connection between Casimir's approach of adding up the modes' zero-point energies \cite{Casimir48} and the Lifshitz formula \cite{Lifshitz56}, provided by the argument principle:
\begin{align}
&\frac{E}{A}=-\sum_{\sigma=s,p}\int_{0}^{\infty}\frac{\textmd{d}K K}{2\pi a^{3}}\sum_{\modeindex}\left[\frac{\Omega_{\sigma, \modeindex}}{2}\right]\bigg{\vert}^{a}_{a\to \infty}
\label{eq:Casimir}\\
&~~~Ê=-\sum_{\sigma=s,p}\int_{0}^{\infty}\frac{\textmd{d}K K}{2\pi a^{3}}\int_{0}^{\infty}{\rm d}\OmegaÊ\frac{\Omega}{2} ~Ê\frac{\partial}{\partial \Omega}  D_\sigma(\Omega, K)~,
\label{eq:Lifshitz}
\\
&D_\sigma(\Omega, K) =  \frac{1}{\pi}\Im \ln\left[1-r^\sigma_{L}(\Omega) r^\sigma_{R}(\Omega)e^{-2\kappa}\right]~.
\end{align}
We recall that the argument principle relates mathematical discontinuities to the modes of the system \cite{VanKampen68, Davies72, Schram73, Intravaia12b}. In fact, the zeros of the argument of the logarithm correspond to the modes of the mixed cavity, according to Eq.\,\eqref{eq:Dispersion}, while its poles give the discrete modes living on the isolated mirrors.
Also, the continua of modes encountered in Sec.\,\ref{sec:continuum} are mathematically described by branch-cuts. 
Hence, the quantity $\partial D_\sigma/\partial \Omega$ fully describes a differential density of states of the electromagnetic field inside the cavity.

Numerical evaluations of the full Casimir potential typically involve an analytic continuation to imaginary frequencies. There, however, modal contributions cannot be separated and we must work at real frequencies to study the mode anatomy of the Casimir potential.

For the following analysis, we express the energy in terms of a correction factor 
$\eta = E / |\mathcal{N} E_C|$
 that compares the cavity's dispersion energy to the absolute value of Casimir's famous energy $ E_C = -{\hbar c A}/({4 \pi \mathcal{N} a^3})$ with $\mathcal{N}=180/\pi^{3}$,
 obtained in the case of perfectly reflecting mirrors \mbox{($ r_{i}^{s,p}=\mp1$)}  \cite{Casimir48}.
A negative (positive) $\eta$ then corresponds  to a binding (antibinding) energy \citep{Henkel04}.
Note that  a factor $\mathcal{N}$ is absorbed into the energy scale to obtain simpler expressions.
In the case of two parallel metallic mirrors, the surface plasmon modes are known to play an important role, despite their evanescent electromagnetic field. There, the plasmonic contribution to the interaction is dominant and attractive at short distances while it is large (larger than the total Casimir force) and repulsive at long distance \cite{Intravaia05, Intravaia07}. However, due to compensations with other modes the total force remains  attractive. 
So far, for configurations using `normal' materials,  repulsion has been predicted in configurations out of thermal equilibrium \cite{Antezza08, Haakh10} or for special geometries \cite{Levin10}.
Calculations involving metamaterials led to several predictions \cite{Henkel05, Rosa08, Rosa08a, Lambrecht08, Pirozhenko08,Yannopapas09} including a repulsive total interaction in some range of distances.
Here, however, a strong magnetic low frequency response is required  \cite{Henkel05, Rosa08, Rosa08a}, ruling out `classical' materials in some cases, according to the Bohr-van Leuwwen theorem \cite{Rahi10}.

\begin{figure}[t!]
\centering
\includegraphics[height=5.5cm]{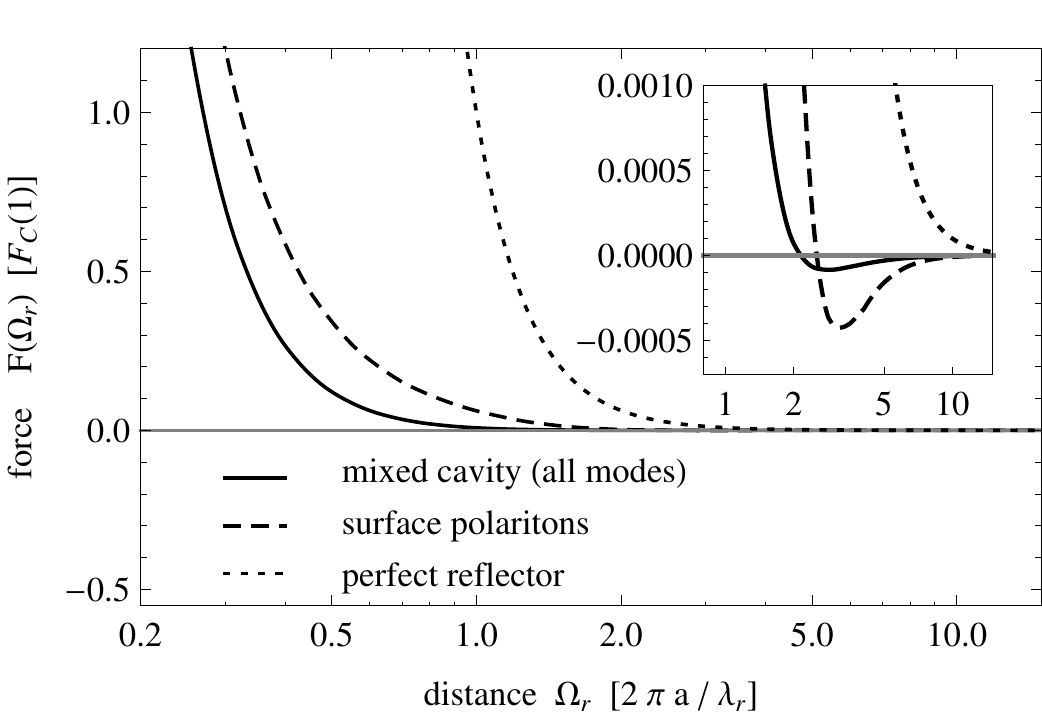}
\hspace{2ex}
\caption{Casimir force in a  mixed metal/magneto-dielectric cavity (${\chi_\epsilon} = 0.25$, ${\chi_\mu} = 0.64$). A positive (negative) value suggests attraction (repulsion). The total force (Lifshitz result) is compared to the contribution of the surface polaritons (see also Sec.\,\ref{sec:discrete_polaritons}). For comparison we also show the force obtained in the limit of a perfectly reflecting cavity, whose value at $\Omega_r = 2 \pi a /\lambda_r=1$ is used as a scale. The inset shows a magnification of the repulsive region that occurs at a cavity length $a \sim \lambda_{r}$. The surface polariton contribution shows a stronger repulsion at slightly larger distances, indicating the important role of these modes in the change of sign of the Casimir force and a partial cancellation with other mode contributions.
\label{fig:force_1}
}

\end{figure}

A numerical evaluation of the Casimir force $F = - \partial E/\partial L$ using the models given in Eq.\,\eqref{eq:model_metamaterial} is given in Fig. \ref{fig:force_1}. It shows indeed that at zero temperature for the parameter used here (${\chi_\epsilon} = 0.25$, ${\chi_\mu} = 0.64$) the Casimir force become repulsive around $2\pi a/\lambda_{r}\sim 2$. In the same figure we also show the contributions from the surface polaritons calculated in the next section, which become repulsive in a very similar range of distances, demonstrating the important role of this contribution in the sign change of the force.

%

\subsection{The role of discrete surface polaritons}
\label{sec:discrete_polaritons}

\begin{figure*}
\centering
a) \includegraphics[height=5.5cm]{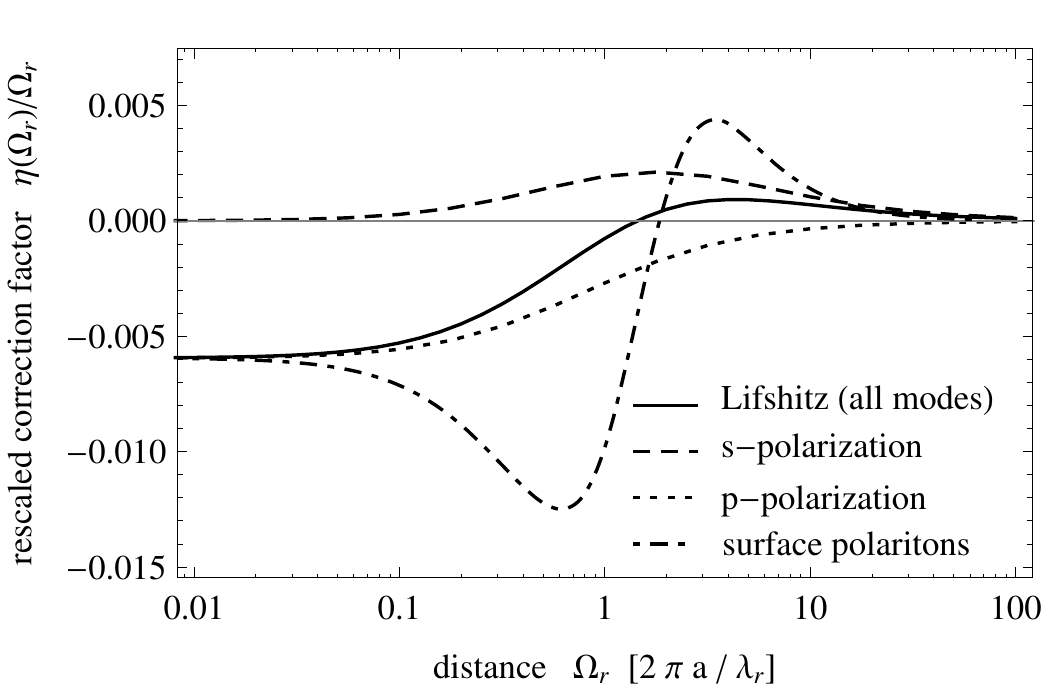} 
b) \includegraphics[height=5.5cm]{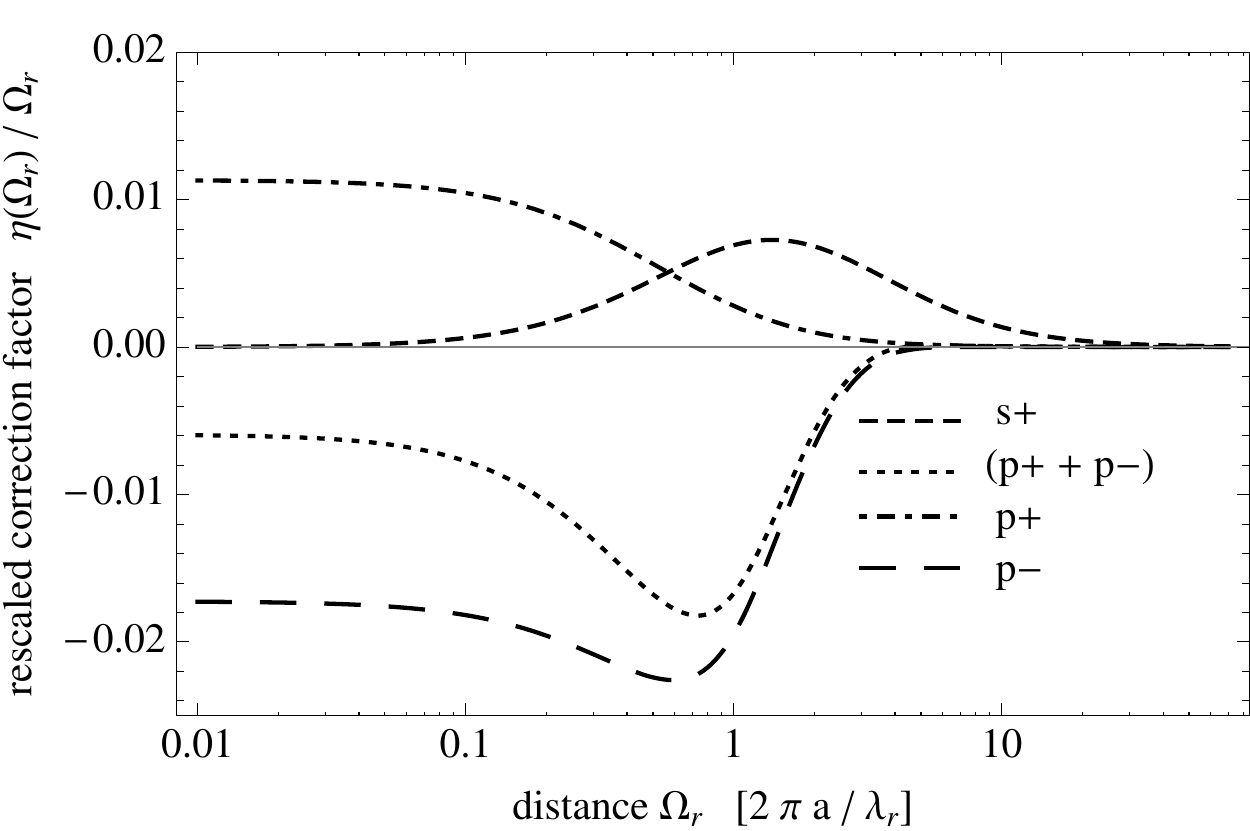}

\caption{Energy correction factor as a function of distance, scaled by one power of the distance.
a) We give the full energy including all modes, obtained using the Lifshitz formula (black solid line), as well as the contributions of all modes of each polarization (dashed and dotted curves) . These are compared with the contributions from the discrete polariton modes. While the \TMabbrev-polarized modes dominates at short separation between the plates, the \TEabbrev -polarized ones becomes more relevant at large separation and dominates at distances $\Omega_r  = 2 \pi a / \lambda_r \sim 2 \cdots 10$ (at the parameters chosen), where the Casimir force becomes repulsive (see also Fig. \ref{fig:force_1}). As in the case of a purely metallic cavity \cite{Intravaia05,Intravaia07,Haakh10}, the polaritonic contribution alone can be much larger than the total contribution, but is partly cancelled as a result of a subtle balance between all modes of the system.
\newline 
b) Anatomic decomposition of the correction factor in terms of the individual surface polariton modes. While both $\eta_{s+}$ and $\eta_{p+}$ are positive (anti binding), $\eta_{p-}$ is negative (binding). Note that the energy contribution of $s+$ is larger in amplitude than the total \TEabbrev -polarized contribution in panel a), but with its maximum at similar distances, indicating the strong role of the \TEabbrev -polariton in creating repulsive regimes.  
}
\label{fig:eta_polaritons}
\end{figure*}


Eq.\,\eqref{eq:Casimir} allows for the direct evaluation of the  energy contribution due to any  polaritonic mode $\Omega_{\modeindex}$:
\begin{equation}
\label{eq:E_diff}
E_{\modeindex}=\eta_{\modeindex} \mathcal{N}|E_{C}|,~\eta_{\modeindex}=\int_{\Gamma_{\modeindex}}\!\! \textmd{d}K K  [\Omega_{\modeindex}(K)-\Omega_{\modeindex0}(K)] ~.
\end{equation}
Here, $\Omega_{\modeindex 0}$ is the corresponding isolated polariton and $\Gamma_{\modeindex}$ indicates the integration contour in the $\Omega$-$K$ plane. While there is a clear interpretation of the integral as a total shift of mode energies, the choice of $\Gamma_{\modeindex}$ may depend on the way in which isolated and coupled modes are identified. 
For a related discussion in the case of a metallic cavity, see Refs.\,\cite{Intravaia07,Haakh10}. Similar to the treatment there, we use closed paths $\Gamma_{\modeindex}$, i.e. we appropriately extend the expression of the surface polariton dispersion relation in such a way that they are both defined over the same range of values of $K$. In doing so we also automatically set the energy contribution to zero for $a\to\infty$ \cite{Intravaia07}.
From the previous analysis we can see that all dispersion relations approach a finite limit as $K\to \infty$.   
For small $K$ values the isolated dispersion relations are continued  along the respective band limits up to the tangent points of the corresponding coupled polariton
\begin{equation}
 \bar\Omega_{\modeindex0}(z) := \begin{cases}
\Omega_{\modeindex0}(z)& z_{\modeindex0} < z < \infty \\
\Omega_{\pm}(z)&z < z_{\modeindex0}~,
\end{cases}
\end{equation}
which corresponds to identifying the cavity mode $\Omega_{\modeindex}(z)$ as the counterpart of the isolated mode $\Omega_{\modeindex0}(z)$ at the same value of $z$. 
It is very useful to change the integration variable according to
%
$K~Ê\textmd{d}K=\textmd{d}z/2+\Omega_{\modeindex}\textmd{d}\Omega_{\modeindex}$
%
so that
\begin{align}
\label{eq: }
\int^{\infty}_{K_{\modeindex}} K\Omega_{\modeindex}(K) \textmd{d}K \nonumber\\
&\hspace{-2cm}=\frac{1}{2}\int^{\infty}_{z_{\modeindex}=z(K_{\modeindex})} \Omega_{\modeindex}(z) \textmd{d}z+\int^{\Omega_{\modeindex}(\infty)}_{\Omega_{\modeindex}(z_{\modeindex})} \Omega^{2}\textmd{d}\Omega \nonumber\\
&\hspace{-2cm}=\frac{1}{2}\int^{\infty}_{z_{\modeindex}} \Omega_{\modeindex}(z) \textmd{d}z+\frac{1}{3}\left[\Omega_{\modeindex}(\infty)^{3}-\Omega_{\modeindex}(z_{\modeindex})^{3}\right].
\end{align}
Our choice of closed contours $\Gamma_n$ allows for further simplification. Since the starting and the end points of the isolated and the coupled polaritons dispersion relations coincide, the last term of the previous expression cancels in the difference of Eq.\,\eqref{eq:E_diff}, so that we finally find
\begin{equation}
\label{eq:paramContr}
\eta_{\modeindex}=\frac{1}{2}\int^{\infty}_{z_{\modeindex}}\left[ \Omega_{\modeindex}(z)- \bar\Omega_{\modeindex0}(z)\right] \textmd{d}z~.
\end{equation}

In Figs.\,\ref{fig:eta_polaritons} and \ref{fig:eta} we plot the functions $\eta_{\modeindex}(\Omega_r)/\Omega_r$ for the different discrete polaritons as a function of the rescaled distance ($\Omega_r=2\pi a/\lambda_{r}$) to emphasize the short-distance behavior.
Similar to the case of two metallic surfaces, the two \TMabbrev -polaritons together give rise to an attractive force which at short distances scales linearly with the distance and dominates the total potential, showing that the \TEabbrev-polarized polariton contribution is sublinear and, hence, subleading in this distance regime \cite{Lambrecht08}.

The physical mechanisms behind the energy shift of the modes are in fact quite different:
In the case of the \TMabbrev-polarized modes, the density of charge associated with the surface states  is coupled by the electromagnetic field \cite{Intravaia07}.
The two isolated polaritons hybridize across the gap between the two surfaces, yielding an antibinding polariton $\Omega_{p+}$ and a binding one $\Omega_{p-}$ so that $\eta_{p+}>0$ and $\eta_{p-}<0$ at all distances.

In contrast, when a metallic and a magneto-dielectric plate are brought close to each other, the \TEabbrev-polarized surface modes 
do not have a counterpart on the metal, and do not undergo this kind of surface-mode coupling and hybridization \cite{Lambrecht08}.
Still, as shown in our previous analysis, the presence of the metallic boundary has an influence on the \TEabbrev -polaritons that leads to the distance dependent dispersion relation $\Omega_{s+}$. The mechanism behind this effect (neglected in previous work \cite{Lambrecht08}) is a distance dependent phase-shift of the electromagnetic field due to the presence of the metal. 
We can immediately understand the antibinding behavior from the shape of the band limits in Fig.\,\ref{fig:bandlimits}: as the end point $z_{p+}$ moves along the band limit $\Omega_+(z)$ from the isolated value $z_{p0+}$ to the coupled value, the whole dispersion curve must necessarily shift towards higher energies. The wave vectors corresponding to the interval $[z_{p+}, z_{p0+}]$ determine the  relevant range of distance.

The resulting curve for the energy correction due to the discrete \TEabbrev-polarized mode is also reported in Fig.\,\ref{fig:eta}.
As one can see the \TEabbrev-polarized polaritonic contribution has an overall positive sign and, although subleading at short distances, it can even turn to be the largest single mode contribution at distances around $a\sim 2...3 \times \lambda_r / 2 \pi$, which is exactly the region where the Casimir force becomes repulsive.

Taking advantage of the analytical expressions for the dispersion relations in the next section, we further investigate the polaritonic contribution providing asymptotic expressions for each contribution at short distances.

\subsection{Short distance behavior}
\label{sec:asymptotes}
\begin{figure}[t!]
\centering
\includegraphics[height=5.5cm]{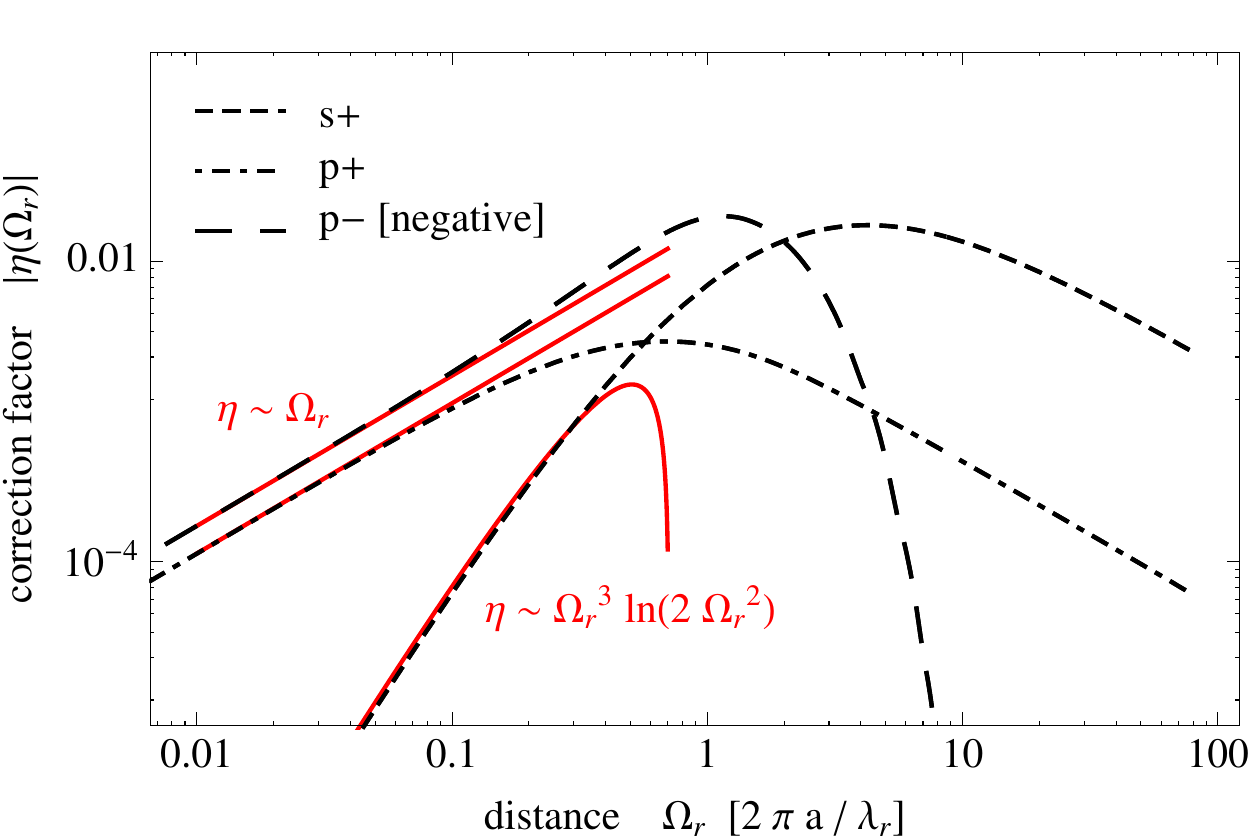}

\caption{(Color online)  Discrete mode contributions to the potential correction factor, doubly logarithmic. The asymptotic expressions are compared with the numerical calculations. As for the metallic cavity case the \TMabbrev -contribution grows linearly with the distance \cite{Intravaia07,Lambrecht08,Haakh10}, while the \TEabbrev -polarization polariton shows an unusual logarithmic behavior.
}
\label{fig:eta}
\end{figure}

At short distances, the contribution from the \TMabbrev -polaritons can be obtained as usual by expanding the dispersion relation found above   \cite{VanKampen68,Genet04,Intravaia05,Intravaia07,Lambrecht08,Haakh10}.  Short distances correspond to large wave vectors or equivalently large $z$ in which limit Eq. \eqref{eq:Dispersion} has the solutions
\begin{align}
\Omega_{p\pm} = \frac{\Omega_r}{2} \sqrt{ (3 + {\chi_\epsilon}) 
\mp \sqrt{4 e^{- 2 \sqrt{z}}  {\chi_\epsilon} + (1 + {\chi_\epsilon})^2
}}~.
\label{eq:omega_pm_nearfieldlimit}
\end{align}
The corresponding limits of the isolated modes are obtained as $z \to \infty$. 
As  expected from the numerical evaluation, Eqs.\,\eqref{eq:paramContr} and \eqref{eq:omega_pm_nearfieldlimit} provide a linear scaling 
$\eta_{p\pm}\approx \beta_{\pm} \Omega_{r} 
$ with the distance,
with  prefactors given by a dimensionless numerical integral.
At our parameters,  the values $\beta_{-}=-0.017$ and $\beta_{+}=0.011$ can be directly read off Fig.\,\ref{fig:eta_polaritons}b).
 This and Fig.\,\ref{fig:eta} show an excellent agreement with the full calculation. 

%
%
From the general formula for the \TEabbrev -polarized polaritonic dispersion relation $\Omega_{s+}$ (see Table \ref{tab:dispersion}), we take the leading order correction for large $z \gg \sqrt{\Omega_r}$ and expand
\begin{align}
C_s 	
& \approx
	 1 + \frac{\Omega_r^2}{2 z} e^{-2 z} ~.
\label{eq:C_limits}
\end{align}
Expanding the dispersion curve gives
%
\begin{align}
\label{eq:Omega_splus_small_L}
\Omega_{s+} / \Omega_r \approx
\frac{\sqrt{2 + {\chi_\mu}}}{\sqrt{2}} 
+ (C_s-1) \frac{{\chi_\mu}}{4 \sqrt{2} \sqrt{2 + {\chi_\mu}}}~.
\end{align}
The first term recovers exactly the isolated polariton $\Omega_{s0+}$ in the same limit ($z\to \infty$).
At short distance the lower bound in Eq.\,\eqref{eq:paramContr} is $z_{s+}\to -\Omega_r^2$, but one can show that the interval $z \in [-\Omega_r^2, \Omega_r^2]$ gives a subleading contribution, so that
%
\begin{align}
\eta_{\rm s +} &= \frac 1 2 \int_{-\Omega_r^2}^{\infty} \textmd{d}z \left[\Omega_{s+}(z) - \Omega_{s0+}(z)\right]\nonumber\\
& \approx   \frac{\Omega_r^3~ {\chi_\mu}}{16 \sqrt{2}\sqrt{2 + {\chi_\mu}}} 
\int_{\Omega_r^2}^\infty \frac{e^{-z}}{z} \textmd{d}z \nonumber\\
&\approx  \frac{\Omega_r^3~ {\chi_\mu}}{16 \sqrt{2}\sqrt{2 + {\chi_\mu}}} 
\ln(2 \Omega_r^2)~.
\end{align}
In the last step we expanded the integral representation of the  $\Gamma$-function for small $\Omega_r$. 
The previous expression indicates a logarithmic contribution for the \TEabbrev-polaritonic contribution and limiting curve given in Fig.\,\ref{fig:eta} shows very good agreement with the numerical integration.
The previous result provides a quantitative explanation of the sublinear scaling encountered in the numerical data (Fig.\,\ref{fig:eta_polaritons}) and traces the different scaling with respect to the \TMabbrev-polarized modes back to the underlying physical mechanism. 

We already showed that the near-field behavior of the total Casimir potential is dominated by the \TMabbrev -polarized polaritons. It is also interesting to consider the two polarizations individually as in Fig.\,\ref{fig:potential} and compare with the corresponding polaritonic contribution. We see that the previous near-field limits provides a good description of the \TMabbrev -polarized contribution, while it overestimates the \TEabbrev -polarized one.

The origin of this difference lies in the fact that in the near field for our configuration, the discrete polaritons are not the only modes that contribute to the Casimir interaction. From the the analysis of the mode spectrum reported in Section \ref{sec:mode_anatomy} the isolated poles are not the only discontinuities of the density of states of our system but we have to complete our analysis also by taking into account the contribution of the  mode continua.

\subsection{Contributions from the bulk continuum}
\label{sec:continuum}

\begin{figure}[t!!]
\centering
\includegraphics[width=\columnwidth]{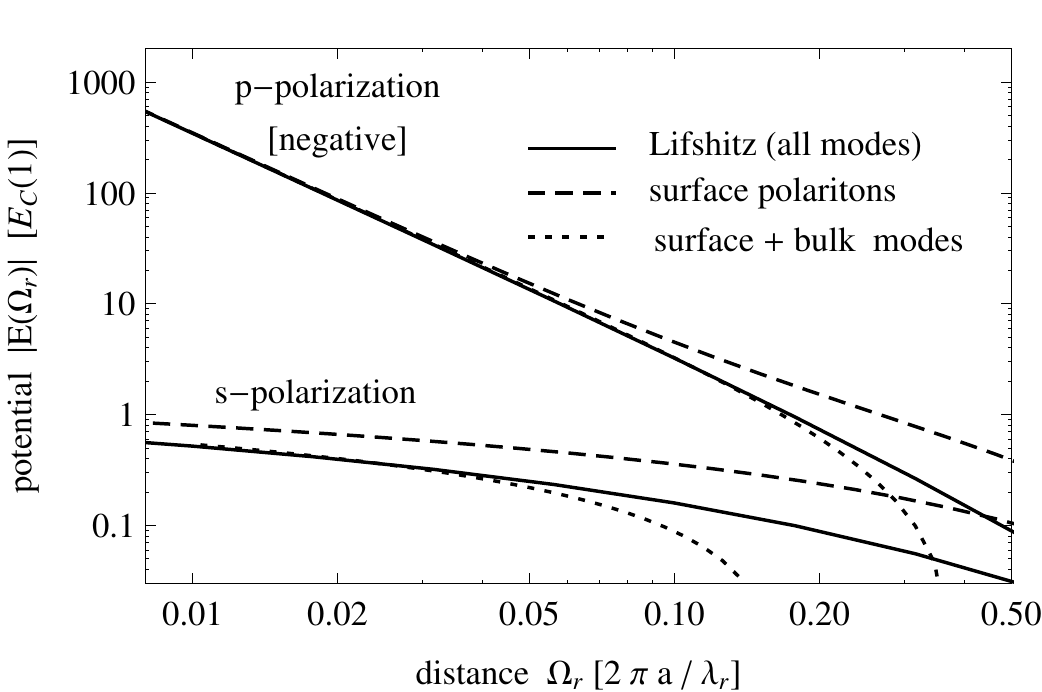}
\caption{Casimir potential energy per polarization as a function of the distance, normalized to the limit of a perfectly reflecting cavity at $\Omega_r = 2 \pi a / \lambda_r = 1$. In \TMabbrev -polarization the contributions of discrete polaritons alone describe well the (linear) short distance behavior of the total \TMabbrev -energy. In \TEabbrev -polarization, the (logarithmic) contribution of the discrete polariton modes overestimates the total \TEabbrev -polarized potential at short distance. This difference is eliminated when we include the contribution of the bulk continuum of \TEabbrev -polarized modes.
}
\label{fig:potential}
\end{figure}

\begin{figure}[t!!]
\centering
\includegraphics[width=\columnwidth]{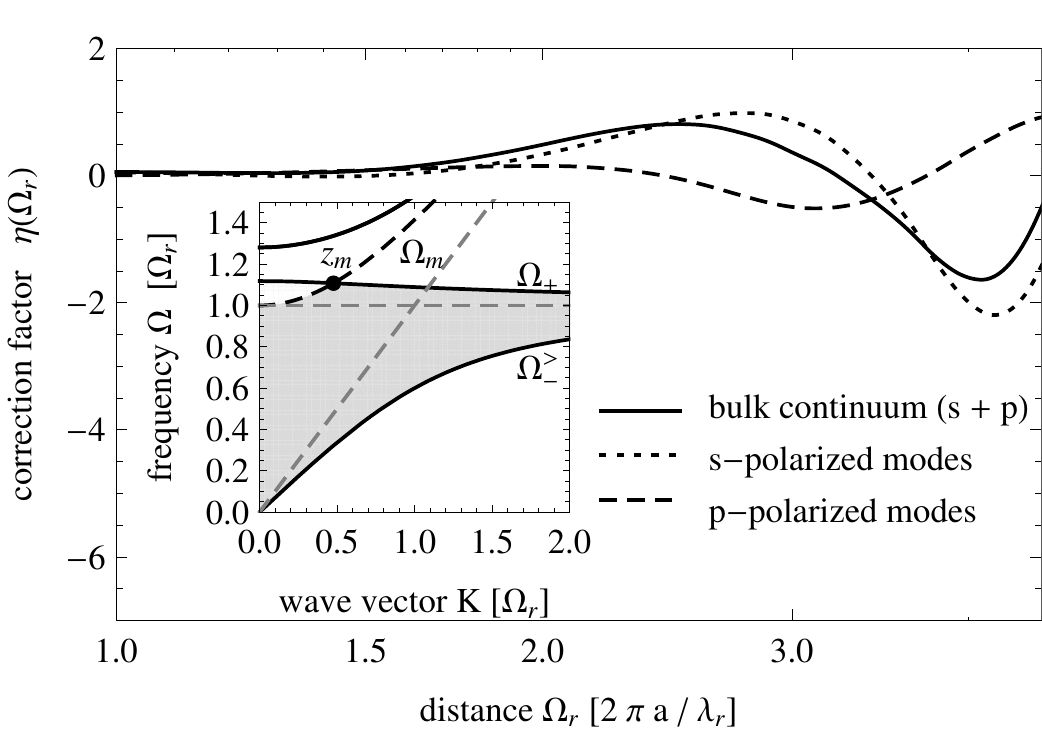} 
\caption{Contribution to the Casimir energy due to the bulk continuum modes,
and contributions of the individual polarizations. In the full Casimir energy, the oscillations at large distances will cancel with other (propagating) mode contributions.
The inset shows the region in the $\Omega$-$K$-plane limited by magneto-dielectric and metal band limits that hosts the bulk modes considered in Sec.\,\ref{sec:continuum}.
,
}
\label{fig:eta_BC}
\end{figure}

We now turn to the bulk continuum of modes limited by the magneto-dielectric and metallic band limits (see Fig.\,\ref{fig:bandlimits} zones ii) and iii\,b), and the inset of Fig.\,\ref{fig:eta_BC}).
These modes are evanescent in the metal, but can propagate in the magneto-dielectric. In the vacuum between the mirrors, these modes can be either evanescent or propagating. They are unique to the mixed metal/magneto-dielectric cavity and have no equivalent in the (lossless) metallic cavity \cite{Intravaia09,Intravaia12b}. 

We calculate their contribution by applying applying Eq.\,\eqref{eq:Lifshitz} to this region of the $\Omega$-$K$-plane and performing a partial integration. In general, a region limited by two curves $\Omega_{a,b}(K)$ gives a contribution to the energy correction factor

\begin{align}
\eta 
 & =
  \sum_{\sigma = s,p}\int_{0}^{\infty}\textmd{d}K K \int_{\Omega_{a}(K)}^{\Omega_{b}(K)}{\rm d}\Omega ~Ê D_\sigma(\Omega, K)\nonumber\\
 &-\sum_{\sigma = s,p}\int_{0}^{\infty}\textmd{d}K K\left[\Omega D_\sigma(\Omega, K)\right]\bigr{\vert}^{ \Omega_{b}(K)}_{\Omega_{a}(K)}. 
\label{eq:Lifshitz2}
\end{align}

In our case, a parametrization is most readily obtained  by considering the full area between the magneto-dielectric band limits [zones  ii) and iii)] and subtracting the contribution due to the fully propagative modes  in region iii\,a).
When calculating the former integral, we further split the regions above and below the light cone [$z>0$]. 
Hence, for the region $\Omega_+, K > \Omega > \Omega_-^>$ we find
\begin{align}
\eta_{>} =&  \frac{1}{2}\sum_{\sigma=s,p} \int_{0}^\infty \textmd{d}z  \biggl[
\int_{\Omega_{-}^>(z)}^{\Omega_+(z)} \textmd{d}\Omega D_\sigma(z, \Omega)\nonumber\\
&-\left[\Omega D_\sigma(z, \Omega)\right]\bigr{\vert}^{\Omega_+(z)}_{\Omega_{-}^>(z)} \biggr]~.
\end{align}
The contribution of the other regions can be considered in a similar way. This gives rise to $\eta_{<}$ for the remaining parts of  ii) and iii) above the light cone, and to $\eta_0$ stemming from region iii\,a).  Finally, the contribution of the continuum modes is given by the sum $\eta_{\rm bulk}=- \eta_0 + \eta_> + \eta_< $.
A numerical evaluation is shown in Fig.\,\ref{fig:eta_BC} together with its decomposition in \TEabbrev- and \TMabbrev-polarization. 

Adding the continuum mode contribution to the surface polariton potential of the previous section, we find good agreement with the full Casimir potential predicted by the Lifshitz formula \eqref{eq:Lifshitz} for \TEabbrev- and \TMabbrev-polarization, as can be seen in Fig.\,\ref{fig:potential}.
As expected from Sec.\,\ref{sec:asymptotes}, the \TMabbrev-polarized mode continuum provides only a minor correction in the near field, where the \TMabbrev-polarized contribution to the Casimir potential is described accurately by the interaction of the discrete polariton modes alone (Fig.\,\ref{fig:potential}). In \TEabbrev -polarization, however, the  mode continuum due to the magneto-dielectric branchcut cannot be neglected in a full description of the Casimir potential.

Finally, it is worth to make some comments on the remaining regions of the $\Omega$-$K$-plane.
The previous set of modes has an exact counterpart in  the region iv\,a), which contains modes that can propagate in the metal but have an evanescent character inside the magneto-dielectric. This band, however, lies well above the plasma edge where the material becomes transparent. As a consequence, it hardly affects the structure of the e.\,m.\ vacuum, leading to very small contributions to the Casimir effect, as could be expected from the lack of a change in the density of states visible in Fig.\,\ref{fig:bandlimits}\,a).
Finally, as we go to larger plate separations, the usual propagating (photonic) modes appear in the region iv).

\section{Summary and Discussion}
\label{sec:discussion}

We have performed a careful analysis of the modes in a mixed cavity formed by a metallic and a magneto-dielectric (meta)material mirror. Using nanostructured composite media consisting of special material like superconductors and/or ferromagnetic materials a large range of values for the magnetic permeability and electric permittivity can be achieved \cite{Du06,Ricci05,Du08,Chen10,Yannopapas09}. 

We first provided analytic forms of the dispersion relations. From there it was possible to calculate the contribution to the Casimir effect due to  discrete polariton modes exactly. Asymptotic evaluations at small distances give good agreement with the full solutions.
At short distances the dominating contribution to the Casimir interaction can be traced back to the \TMabbrev -polarized surface modes that hybridize accross the cavity, similar to the known results from entirely metallic cavities and in qualitative agreement with previous studies using different models for the  response of a  metamaterial \cite{Henkel05,Lambrecht08}.
We find that the \TEabbrev -polarized surface modes provide a large contribution to the repulsive behavior of the Casimir force for distance of the order of the resonance wavelength showing the importance of surface polaritons in the mechanism of repulsion. We attribute the distance dependence of this component to a phase shift the single \TEabbrev -polariton living on the magneto-dielectric mirror acquires in the presence of the second surface, a mechanism quite different from the mode hybridization at work in the \TMabbrev -polarization. Besides, the effects of the binding and anti-binding \TMabbrev-polarized modes partially cancel each other out, which does of course not occur for the single \TEabbrev-polariton.

 It is worth noticing that the frequencies involved in the description of surface-polaritons are of the order of the resonance frequency and therefore the repulsive contribution due to these modes is  sensitive only to the form of $\epsilon(\omega)$ and $\mu(\omega)$ for $\omega\sim \omega_{r}$. At the fist glance, this seems to be in contradiction with the claim that the low frequency behavior of the material's magnetic response is relevant for a repulsive Casimir force \cite{Rosa08a,Rosa08}. This issue can be resolved by recalling that repulsion occurs at distances, where other (propagating photonic) modes becomes relevant. These modes lead to an attraction that works against the polaritonic repulsion.
The balance of all contribution leads to a total Casimir force that features a repulsive regime that is much weaker but always occurs at plate separations of the order of the material resonance wavelength.
A modification of the magnetic response at low frequencies hardly affects the repulsive behavior of the polaritonic modes, while it may have a strong impact on the capability of the photonic ones to provide a compensating attractive force. Once again, this unterlines the subtle balance of different sets of modes that is characteristic of the Casimir effect.
While the repulsion is known to be frail in thermal equilibrium at nonzero temperature, we may speculate that nonequilibrium excitations of the surface polaritons may provide a means of maintaining repulsion even at higher temperatures, similar to the findings of Ref.\,\citep{Haakh10}.

In addition to the previous findings we have also shown that the bulk continuum of \TEabbrev -polarized modes inside the magneto-dielectric band provides a relevant contribution to the near-field Casimir potential - an effect that has no counterpart in a purely metallic cavity.

Finally, the closed forms of the surface mode dispersion relations may be a handy toolbox for other calculations of recent interest, such as the nonradiative heat transfer between two cavity walls at different temperatures, which could be modified via the additional degree of control provided by a taylored magnetic response of the cavity wall.

\section*{Aknowledgments}
We are indebted to C. Henkel and D.A.R. Dalvit for helpful discussions and comments on the manuscript. Partial financial support by the Potsdam Graduate School, by the German-Israeli-Foundation for Scientific Research and Development (GIF), and by Los Alamos National Laboratory is gratefully acknowledged.

\appendix

\section{Band limits and branch cuts }
\label{app:band_limit}
In the following Appendices, we give  details of the calculations leading to the dispersion curves and band limits given in Tab.\,\ref{tab:dispersion} and discuss their behavior in some important limits.
For the present analysis, it is convenient to use the following identities for the response functions
%
\begin{align}
\label{eq:ImportantRelation1}
\epsilon_{L}\epsilon_{R} &=(1+{\chi_\epsilon})\epsilon_{L}-{\chi_\epsilon}\quad\\
\label{eq:ImportantRelation2}
\epsilon_{L} \mu_R&=(1+{\chi_\mu})\epsilon_{L}-{\chi_\mu}~,
\end{align}
%
%
where we use the particular form of $\epsilon_{L}=1-\Omega_{r}^{2}/\Omega^{2}$.
We can write the propagation constant $\kappa_{R}$ inside the magneto-dielectric 
in the form
\begin{equation}
\label{eq:ImportantRelation3}
\epsilon^{2}_{L}\kappa_{R}^{2}=\epsilon^{2}_{L}z+\NewDelta \epsilon_{L}-\Newdelta ~,
\end{equation}
where  $z=K^{2}-\Omega^{2}$ as before and where we abbreviate
$\NewDelta =\Omega^{2}_{r}({\chi_\mu}\,{\chi_\epsilon}+{\chi_\mu}+{\chi_\epsilon})$ and $\quad  \Newdelta =\Omega^{2}_{r}{\chi_\mu}\,{\chi_\epsilon}$.
%
%
%
The magneto-dielectric can behave as a left-handed material if the propagation constant $\kappa_R$ changes its sign. This occurs, if
$\kappa_{R}=0 \Rightarrow \epsilon^{2}_{L}z+\NewDelta \epsilon_{L}-\Newdelta =0$.
%
This second order equation has as its solutions the two relations $\epsilon_L= \epsilon_{\pm}(z)$ given in Table \ref{tab:dispersion} and the dispersion relations follow by solving for the frequency
[cf. Eq.\,\eqref{eq:omega_from_epsilon}].

The value of 
$\Omega_+(K=0)$ is obtained by setting $z = -\Omega^2$ in the previous expression, so that 
\begin{align}
\Omega_+({K=0}) &= \nonumber
\textstyle \sqrt{\Omega_{r}^{2}+\frac{\NewDelta -\Newdelta }{2}-\sqrt{\left(\frac{\NewDelta -\Newdelta }{2}\right)^{2}-\Omega_{r}^{2}\Newdelta }}~,
\\
&=\Omega_r \sqrt{1 + \chi_\mu}
\label{eq:Omega_plusK0}
\end{align}
The solution $\Omega_-$ has two branches for $z>0$ and $z<0$. At zero momentum, the former vanishes $\Omega_{-}^{>}(K=0)=0$, while the latter takes the value 
\begin{align}
\label{eq:Omega_minusK0}
\Omega_{-}^{<}({K=0})=
\Omega_r \sqrt{1 + \chi_\epsilon}
\end{align}
The respective starting points are then given by \mbox{$z_\pm = - [\Omega_{\pm}(K=0)]^2$}.
Note that the band limits are monotonous functions over the whole physical range of momenta.

Finally, the band limit arising from the propagation constant $\kappa_L$ in the metal is simply
\begin{align}
\Omega_m = \sqrt{K^2 + \Omega_r^2}~.
\end{align}
Since this implies a constant value of $z= z_m = - \Omega_r^2$, the metal band limit $\Omega_m$ and the magneto-dielectric's one cross at an energy given by $\Omega_+(z_m)$.
\section{Surface polariton dispersion relations}
\label{app:SPDR}
\paragraph{Isolated modes.} The isolated surface polaritons living on the interface of each medium with vacuum are given by the solutions of Eq.\,\eqref{eq:isolated}.
For the metallic mirror we find the usual \TMabbrev -polarized surface plasmon dispersion relation  $\Omega_{p0-}$,
%
%
 which goes to zero as $K\to0$ and tends to $\Omega_{r}/\sqrt{2}$ for $K\to \infty$ \cite{Intravaia05,Intravaia07,Haakh10}.
In contrast, the magneto-dielectric mirror features two unconventional polaritons, one in each polarization. 
 We multiply the remaining Eqs.\,\eqref{eq:isolated} by $\epsilon_L$, take the square and use   relations \eqref{eq:ImportantRelation1}-\eqref{eq:ImportantRelation3} to find the following second order equations in $\epsilon_L$:
\begin{align}
\epsilon^{2}_{L}z+\NewDelta \epsilon_{L}-\Newdelta &= [(1+{\chi_{j}})\epsilon_{L}-{\chi_{j}}]^2 z~, \quad j = \epsilon, \mu.
\end{align}
The  physical solutions $\epsilon_{p0+}, \epsilon_{s0+}$ are reported in Table \ref{tab:dispersion}. 
Both curves stop where the band limit meets the light cone, as discussed before, corresponding to a frequency %
$\Omega_+[z=0]
=\Omega_{r}\sqrt{
\frac{\NewDelta }{
\NewDelta -\Newdelta }
}~.$
%
 However, they approach different limits as $K\to\infty$
\begin{align}
&\epsilon_{s0+}\approx\frac{\tilde{\mu}-1}{\tilde{\mu}+1}
\Rightarrow \Omega_{s0+}\to\Omega_{r}\sqrt{\frac{\tilde{\mu}+1}{2}}\\
&\epsilon_{p0+}\approx\frac{\tilde{\epsilon}-1}{\tilde{\epsilon}+1}
\Rightarrow \Omega_{p0+}\to\Omega_{r}\sqrt{\frac{\tilde{\epsilon}+1}{2}}~.
\label{eq:p0plus_farfield}
\end{align}
Here $\tilde \mu = 1 + \chi_{\mu}$ and $\tilde \epsilon = 1 + \chi_{\epsilon}$ are the static magnetic and electric response of the magneto-dielectric.

\paragraph{Coupled \TEabbrev-polarized polariton mode.} In \TEabbrev -polarization, the condition for a cavity mode is given by the first of Eq.\,\eqref{eq:coupled}. We take the square and get the following second order polynomial equation
\begin{equation}
\epsilon^{2}_{L}\frac{1-\tilde{\mu}^{2}C_s^{2}}{{\chi_\mu}}+2\left(\frac{\NewDelta }{2z {\chi_\mu}}+\tilde{\mu}C_s^{2}\right)\epsilon_{L}-\left(\frac{\NewDelta }{z {\chi_\mu}}+{\chi_\mu}C_s^{2}\right)=0.
\end{equation}
Only one of the two solutions satisfies the original equation ($\epsilon_{s+}$, given in Table \ref{tab:dispersion}) whereas the second is a spurious one introduced when taking the square.

The asymptotic form of $\Omega_{s+}$ in the short distance limit was already given in Eq.\,\eqref{eq:Omega_splus_small_L}, showing that the leading term tends to the corresponding isolated surface polariton in the limit $K\to\infty$.
In the opposite limit of large plate separations $a \to \infty$, since $C_s\approx1$,  we recover the isolated polariton.
The coupled \TEabbrev -polarized polariton mode crosses the light cone as $z\to0$. In this limit, we have $z C_s^{2}\to \zeta = (1 + 1/\Omega_r)^{-2} \to 0, 1$ at short and large distances, respectively. Hence we find
\begin{align}
\label{eq:splus_lightcone}
&\epsilon_{s+}(z=0)=\\
& ~~Ê\frac{\left(\frac{\Newdelta }{{\chi_\mu}}+\zeta{\chi_\mu}  \right)}{\left(\zeta \tilde{\mu}+\frac{\NewDelta }{2{\chi_\mu}}\right)
+\sqrt{\left( \zeta \tilde{\mu} +\frac{\NewDelta }{2{\chi_\mu}}\right)^{2}-\frac{\zeta \tilde{\mu}^{2}}{{\chi_\mu}}
\left(\frac{\Newdelta }{{\chi_\mu}}+{\zeta \chi_\mu}\right)}}~.\nonumber
\end{align}
It is worth noticing, that this value is nonzero and finite.
%
A final value of interest is the tangent point with the band limit $\Omega_+$, described by $z_{s+}$. 
Since the band limit is a solution of $\kappa_{R}=0$,  the tangent point can be obtained from Eq.\,\eqref{eq:coupled} as the first zero of $C_s=0$, i.e.
\begin{align}
\label{eq:def_zsplus}
\sqrt{\frac{z}{z+\Omega_{r}^{2}}}&=-\coth(\sqrt{z}).
\end{align}
The solution can  therefore be found in the interval $-\frac{\pi^{2}}{4}\le z_{s+}\le 0$. 
In this point, we find
\begin{align}
\label{eq:exactTEz+}
\epsilon_{s+}(z=z_{s+})
&\xrightarrow{\Omega_{r}\gg1}
\frac{\Newdelta }{\NewDelta } =\epsilon_{s0}(z=0),
\end{align}
where we used that  $z_{s+}\approx -\frac{\pi^{2}}{4}$ as  $\Omega_{r}\gg1$ (large distance limit) and
points out how the coupled polariton approaches the isolated one, with the stopping point given by the intersection of the band limit and the light cone.
In the opposite limit of small plate separations, we find  that  $z_{s+}\approx -\Omega_{r}^{2} = z_m$, coinciding with the  metal band limit. As an important implication, the coupled polariton cannot penetrate the metallic bulk continuum.

%

\paragraph{Coupled \TMabbrev-polarized polariton mode.} In \TMabbrev-polarization we take the square of the second condition provided by Eq.\,\eqref{eq:coupled} and obtain the  fourth-order equation
\begin{multline}
\label{eq:TM-Main}
[(1+{\chi_\epsilon})\epsilon_{L}-{\chi_\epsilon}]^{2}(1+\epsilon_{L}T_{0}T_{s})^{2}=\\
(\epsilon_{L}T_{s}+T_{0})^{2}
\left(\epsilon^{2}_{L}+\frac{\NewDelta }{z}\epsilon_{L}-\frac{\Newdelta }{z}\right),
\end{multline}
which again has the two relevant solutions $\epsilon_{p\pm}$ (the remaining ones are spurious) that can be calculated analytically or using computer algebra systems. Their form is, however, not very instructive and we do not give them explicitly.
Some important limits can be extracted directly and identify the physical solutions.
The limit $z\to \infty$ is quite simple to be calculated
\begin{equation}
\label{eq: }
[(\tilde{\epsilon}\,\epsilon_{L}-{\chi_\epsilon})^{2}+\epsilon^{2}_{L}](1+\epsilon_{L})=0 
\end{equation}
which is solved by
\begin{align}
\label{eq:Omega_p_plus_asympt}
&\epsilon_{p+}\approx\frac{\tilde{\epsilon}-1}{\tilde{\epsilon}+1}\Rightarrow \Omega_{p+}\to\Omega_{r}\sqrt{\frac{\tilde{\epsilon}+1}{2}}\\
\label{eq:Omega_p_minus_asympt}
& \epsilon_{p-}\approx-1 \Rightarrow \Omega_{p-}\to\frac{\Omega_{r}}{\sqrt{2}}~.
\end{align}
These are the same limits found from the corresponding isolated polaritons.
In the same way, the value on the light cone ($z=0$) can be obtained by solving
\begin{equation}
[(1+{\chi_\epsilon})\epsilon_{L}-{\chi_\epsilon}]^{2}=
\left(\frac{\epsilon_{L}}{\Omega_{r}}+1\right)^{2}
(\NewDelta \epsilon_{L}-\Newdelta ).
\end{equation}
%

Let us now investigate the behavior of the coupled \TMabbrev-polarized polaritons as they approach the band limits. This happens, when the right hand side of Eq.\,\eqref{eq:TM-Main} vanishes ($\kappa_{R}=0$). 
The only compatible solution provides the condition
\begin{equation}
(1+\epsilon_{L} T_{0}T_{s})=0\Rightarrow \epsilon_{L} =-\coth(\sqrt{z})\sqrt{\frac{z+\Omega_{r}^{2}}{z}}~.
\end{equation}
We  must still fix which of the two band limit curves is to coincide with $\epsilon_L$.
Considering  $\epsilon_{L} = \epsilon_-$ first, gives the point where the binding polariton  ($\Omega_{p-}$) results  tangent to $\Omega_{>}^{-}$ in the evanescent region. We have therefore in the evanescent zone ($z>0$)
\begin{equation}
\label{eq:def_zpminus}
\frac{\NewDelta }{2}+\sqrt{\left(\frac{\NewDelta }{2}\right)^{2}+\Newdelta z} =\sqrt{z}\coth(\sqrt{z})\sqrt{z+\Omega_{r}^{2}}.
\end{equation}
We will call the corresponding solution $z_{p-}$. Note that $z_{p-}$ grows with $\Omega_{r}$ and that the case $z_{p-}=0$ is compatible only with
\begin{equation}
\Omega_{r}\ge ({\chi_\epsilon}+{\chi_\mu}+{\chi_\epsilon}\,{\chi_\mu})^{-1}.
\end{equation}
This means that the coupled entirely evanescent \TMabbrev-polarized surface mode meets the magneto-dielectric band limit in a tangent point a certain values of $\Omega_{r}$ (see also App.\,\ref{app:tangent}).
Finally, turning to $\epsilon_L = \epsilon_{+}$, we get the point in the propagating region where the antibinding polariton $\Omega_{p+}$ is tangent to the band limit $\Omega_+$.
This occurs in the propagating region and we have (solution for $-y^{2}=z<0$)
\begin{align}
\label{eq:def_zpplus}
\frac{\Newdelta }{\frac{\NewDelta }{2} + \sqrt{\left(\frac{\NewDelta }{2}\right)^{2}-\Newdelta y^{2}}}=\frac{\cot(y)}{y}\sqrt{\Omega_{r}^{2}-y^{2}}
\end{align}
Here,  the corresponding solution $z_{p+}=-y_{+}^{2}$ is limited by $y_{+}\le \min\{\Omega_{r},\pi/2\}$. In the limit $\Omega_{r}\ll1$ the solution is $y\sim \Omega_{r}$.

\section{Existence of the solutions and the tangent condition}
\label{app:tangent}
%

To simplify our calculation, we have assumed so far that the plasma frequency of the metallic plate coincides with both the magnetic and electric resonance frequency of the Lorentz-Drude model ($\Omega_{p}\equiv\Omega_{r}$).
In general, these assumptions may be relaxed and we can still parametrize the frequency in terms of the $\epsilon_L$ of the metallic plate (equivalent to Eq.\,\eqref{eq:omega_from_epsilon}) or solve all the equations directly for the frequency.
This leads to an increased mathematical complexity, but adds little to the physical phenomena. 

As an example, we discuss here the impact of a resonance frequency $\Omega_p \ne \Omega_r$ starting from the tangent condition and the behavior of the surface polaritons.
We will consider only  the \TEabbrev -polarization but the arguments for the \TMabbrev -polarization go along the same line.
In general, the upper limit of magneto-dielectric band limit $\Omega_+$ may or may not cross the dispersion relation of the metallic bulk plasmons $\Omega_{m}(K)=\sqrt{K^{2} + \Omega^{2}_{p}}$ at $z=-\Omega_{r}^{2}$. 
Note that $z_{s+}$ still depends exclusively  on the parameters of the metallic plate, namely the plasma frequency. If the band limits  $\Omega_+$ and $\Omega_{m}$ cross, they provide a lower limit to the  tangent point $z_{s+}<0$.
It is, however, 
legitimate to wonder whether the coupled plasmon exists behind the tangent point, i.e. if it is possible to find a solution $\Omega_{s+}(z)$ for $z<z_{s+}$.

Consider again the  equation $\mu_{R}\sqrt{z}C_s =-\kappa_{R}$,
that governs the \TEabbrev -polarized mode [Eq.\,\eqref{eq:coupled}].
We have that
\begin{equation}
\sqrt{z}C_s=
\begin{cases}
>0,  &0>z>z_{s+}\\
\to \Omega_{p}/(1+\Omega_{p})>0,  &z\to 0\\
<0, &-\Omega^{2}_{p}<z<z_{s+}\\
\text{complex valued}, &z<-\Omega^{2}_{p}~.
\end{cases}
\end{equation}
The function $\mu_{R}$ is positive for $\Omega>\Omega_{r}\sqrt{1+{\chi_\mu}}> \Omega_{+}$ (or $\Omega < \Omega_r$), where Eq.\,\eqref{eq:Omega_plusK0} provides the last inequality. The propagation constant $\kappa_{R}$ is real and positive in the Reststrahlenband, i.e. zone iv) of Fig.\,\ref{fig:bandlimits}, and it is imaginary within the magneto-dielectric bulk continuum [zones ii) and iii)].
In particular, $\mathrm{Im}(\kappa_{R})>0$ only if $\Omega_r < \Omega < \Omega_{+}$.  This indicates that solutions with $z<z_{+}$ may exist only for frequencies far above $\Omega_{+}$ and would require a discontinuity for $z>z_{+}$. These solutions do not exist and hence the coupled polariton stops when it is tangent to the dispersion relation $\Omega_+$.
The tangent condition suggests that the dispersion relation can be continued along $\Omega_+$ for all $z<z_{+}$. If $\Omega_{p}\le \Omega_+(K=0)$ there is a  tangent point before $\Omega_+$ crosses the $K=0$ axis and before the point where $\Omega_{m}$ crosses $\Omega_+$. This is essentially due to the fact that $\sqrt{z}C_s$ takes general complex values for $z<-\Omega^{2}_{p}$ so that the equation for the polariton cannot be satisfied.
If $\Omega_{p}> \Omega_+(K=0)$, there may be a coupled polariton mode that crosses the $K=0$ axis without being tangent to the band limit $\Omega_+(K)$. This case is not considered in this paper.


\bibliographystyle{prsty}


\end{document}